\documentclass[letterpaper,showpacs,preprintnumbers,amsmath,amssymb,nofootinbib,onecolumn,superscriptaddress,notitlepage]{revtex4-1}
\usepackage{amssymb}
\usepackage{graphicx}
\usepackage{amsmath}
\usepackage{multirow}
\usepackage{verbatim}
\usepackage{rotating}
\usepackage[usenames,dvipsnames]{color}
\usepackage[bookmarks=true, bookmarksnumbered=true, colorlinks=true,
  urlcolor=darkblue, citecolor=darkgreen, linkcolor=darkred,
  breaklinks=true]{hyperref} 
\usepackage{xspace}

\definecolor{darkblue}{rgb}{0,0,0.5}
\definecolor{darkgreen}{rgb}{0.1,0,0.3}
\definecolor{darkred}{rgb}{0.6,0,0}

\newcommand{\nc}{\newcommand}
\nc{\like}{\mathcal{L}}
\nc{\Emin}{{E_{\rm min}}}
\nc{\Emax}{{E_{\rm max}}}
\nc{\pdf}{\mathcal{P}}
\begin{document}

\preprint{IFIC/16-23, IPPP/16/37}


\title{Analysis of the 4-year IceCube high-energy starting events} 

\author{Aaron C. Vincent}
\email{aaron.vincent@durham.ac.uk}
\affiliation{Institute for Particle Physics Phenomenology (IPPP),\\ Department of Physics, Durham University, Durham DH1 3LE, UK.}
\author{Sergio Palomares-Ruiz}
\email{sergiopr@ific.uv.es}
\affiliation{Instituto de F\'{\i}sica Corpuscular (IFIC)$,$
 CSIC-Universitat de Val\`encia$,$ \\  
 Apartado de Correos 22085$,$ E-46071 Valencia$,$ Spain}
\author{Olga Mena}
\email{omena@ific.uv.es}
\affiliation{Instituto de F\'{\i}sica Corpuscular (IFIC)$,$
 CSIC-Universitat de Val\`encia$,$ \\  
 Apartado de Correos 22085$,$ E-46071 Valencia$,$ Spain}

\begin{abstract}
\vspace{1cm}
After four years of data taking, the IceCube neutrino telescope has detected 54 high-energy starting events (HESE, or contained-vertex events) with deposited energies above 20~TeV. They represent the first ever detection of high-energy extraterrestrial neutrinos and therefore, the first step in neutrino astronomy. In order to study the energy, flavor and isotropy of the astrophysical neutrino flux arriving at Earth, we perform different analyses of two different deposited energy intervals, [10~TeV $-$ 10~PeV] and [60~TeV $-$ 10~PeV]. We first consider an isotropic unbroken power-law spectrum and constrain its shape, normalization and flavor composition. Our results are in agreement with the preliminary IceCube results, although we obtain a slightly softer spectrum. We also find that current data are not sensitive to a possible neutrino-antineutrino asymmetry in the astrophysical flux. Then, we show that although a two-component power-law model leads to a slightly better fit, it does not represent a significant improvement with respect to a single power-law flux. Finally, we analyze the possible existence of a North-South asymmetry, hinted at by the combination of the HESE sample with the through-going muon data. If only using HESE data, the scarce statistics from the northern hemisphere does not allow us to reach any conclusive answer, which indicates that the HESE sample alone is not driving the potential North-South asymmetry. 

 \end{abstract}


\maketitle

\tableofcontents

\section{Introduction}
\label{sec:intro}

After a few  years of operation, the IceCube Neutrino Observatory has provided compelling evidence for the existence of high-energy astrophysical neutrinos~\cite{Aartsen:2013bka, Aartsen:2013jdh, Aartsen:2014gkd, Aartsen:2015zva}. While their extraterrestrial nature is established at more than 6$\sigma$ confidence level (C.L.)~\cite{Aartsen:2015zva}, their precise origin  is still unknown. The signal remains compatible with an isotropic flux~\cite{Aartsen:2013bka, Aartsen:2013jdh, Aartsen:2014gkd, Aartsen:2015zva} and flavor composition analyses~\cite{Mena:2014sja, Palomares-Ruiz:2014zra, Palomares-Ruiz:2015mka, Vincent:2015woa, Watanabe:2014qua, Palladino:2015zua, Aartsen:2015ivb, Bustamante:2015waa, Shoemaker:2015qul, Arguelles:2015dca, Xu:2014via, Aartsen:2015knd}, energy spectrum fits~\cite{Aartsen:2013bka, Aartsen:2013jdh, Aartsen:2014gkd, Aartsen:2015zva, Chen:2013dza, Winter:2014pya, Aartsen:2014muf, Watanabe:2014qua, Kalashev:2014vra, Palomares-Ruiz:2015mka, Vincent:2015woa}, studies of the arrival directions and possible correlations with sources~\cite{Aartsen:2013jdh, Aartsen:2014gkd, Aartsen:2015zva, Aartsen:2015knd, Razzaque:2013uoa,  Bai:2013nga, Ahlers:2013xia, Lunardini:2013gva, Padovani:2014bha, Ahlers:2014ioa, Bai:2014kba, Esmaili:2014rma, Moharana:2015nxa, Emig:2015dma, Santander:2015kka, Neronov:2015osa, Miranda:2015ema, Aartsen:2015dml, Padovani:2016wwn, Neronov:2016bnp} are all unable to single out the neutrino production mechanism or the cosmic source population. This discovery has triggered many different studies proposing different sources (see, e.g., Refs.~\cite{Laha:2013lka, Anchordoqui:2013dnh, Murase:2014tsa} for a general discussion), from standard cosmic accelerators, such as active galactic nuclei~\cite{Kalashev:2013vba, Stecker:2013fxa, Murase:2014foa, Tjus:2014dna, Krauss:2014tna, Dermer:2014vaa, Tavecchio:2014iza, Sahu:2014fua, Kalashev:2014vya, Tavecchio:2014eia, Kimura:2014jba, Petropoulou:2015upa}, star-forming galaxies~\cite{Murase:2013rfa, Tamborra:2014xia, Anchordoqui:2014yva, Chang:2014hua, Bartos:2015xpa}, gamma-ray bursts~\cite{Cholis:2012kq, Liu:2012pf, Murase:2013ffa, Razzaque:2013dsa, Fraija:2013cha, Petropoulou:2014lja, Dado:2014mea, Razzaque:2014ola, Tamborra:2015qza, Tamborra:2015fzv, Senno:2015tsn}, hypernova and supernova remnants~\cite{Fox:2013oza, Murase:2013ffa, Liu:2013wia, Bhattacharya:2014sta, Chakraborty:2015sta, Tamborra:2015fzv, Senno:2015tsn}, the galactic halo~\cite{Taylor:2014hya}, galaxy clusters~\cite{Zandanel:2014pva}, microquasars~\cite{Anchordoqui:2014rca},  neutron stars mergers~\cite{Gao:2013rxa}, tidal disruption events of supermassive black holes~\cite{Wang:2015mmh}, and from a more general perspective, relating the neutrino flux to the cosmic-ray spectrum~\cite{Kistler:2013my, Gupta:2013xfa, Anchordoqui:2013qsi, Neronov:2013lza, Liu:2013wia, Joshi:2013aua, Katz:2013ooa, Fang:2014uja, Kachelriess:2014oma, Dado:2014mea, Winter:2014pya, Anchordoqui:2014pca}, to other more exotic possibilities~\cite{Borriello:2013ala, Feldstein:2013kka, Barger:2013pla, Esmaili:2013gha, Bai:2013nga, Ema:2013nda, Akay:2014tga, Alikhanov:2014uja, Bhattacharya:2014vwa, Anchordoqui:2014hua, Ioka:2014kca, Ng:2014pca, Zavala:2014dla, Stecker:2014xja, Higaki:2014dwa, Ibe:2014pja, Bhattacharya:2014yha, Ema:2014ufa, Blum:2014ewa, Rott:2014kfa, Araki:2014ona, Akay:2014qka, Aeikens:2014yga, Illana:2014bda, Esmaili:2014rma, Cherry:2014xra, Fong:2014bsa, Stecker:2014oxa, Guo:2014laa, Daikoku:2015vsa, Kopp:2015bfa, Murase:2015gea, Davis:2015rza, Esmaili:2015xpa, Roland:2015yoa, Anchordoqui:2015lqa, Boucenna:2015tra, DiFranzo:2015qea, Ko:2015nma, Araki:2015mya, Chianese:2016opp, Dev:2016uxj, Ema:2016zzu, Fiorentin:2016avj, Dev:2016qbd, DiBari:2016guw}.

Multi-messenger analyses, correlating high-energy neutrino fluxes and their $\gamma$-ray counterparts, may be able to distinguish between hadronuclear ($pp$) and photohadronic ($p\gamma$) cosmic-ray processes~\cite{Murase:2013rfa, Ahlers:2013xia, Chang:2014hua, Wang:2014jca, Zandanel:2014pva, Murase:2015xka, Ando:2015bva, Bechtol:2015uqb, Kistler:2015ywn, Chang:2016ljk}.  Indeed, if $pp$ sources were to explain the IceCube event spectrum, a neutrino flux softer than $E_\nu^{-2.2}$ (or a low-energy cutoff) would be required in order to avoid overshooting the $\gamma$-ray cascade limit~\cite{Murase:2013rfa}. This is further constrained by cross correlating $\gamma$-ray sources with galaxy catalogs~\cite{Ando:2015bva}, although it can be modestly mitigated if synchrotron losses at the source are important~\cite{Chang:2014hua}. Another powerful tool to single out the production mechanism of high-energy neutrinos is the study of their flavor composition~\cite{Rachen:1998fd, Athar:2000yw, Beacom:2003zg, Anchordoqui:2003vc, Kashti:2005qa, Serpico:2005bs, Kachelriess:2006fi, Rodejohann:2006qq, Mena:2006eq, Lipari:2007su, Pakvasa:2007dc, Esmaili:2009dz, Choubey:2009jq, Lai:2009ke, Hummer:2010ai, Meloni:2012nk, Fu:2012zr, Vissani:2013iga, Chatterjee:2013tza, Xu:2014via, Fu:2014isa}. As these neutrinos travel over cosmic distances, oscillation probabilities are averaged out during propagation~\cite{Learned:1994wg} and, in the standard production mechanism via $pp$ interactions, after the decays of secondary muons and pions, the expected flavor ratio at the Earth is $(\nu_e:\nu_\mu:\nu_\tau)_\oplus = (1:1:1)_\oplus$ for both neutrinos and antineutrinos. On the other hand, $p\gamma$ sources mainly produce positively charged pions and thus a different flavor ratio is obtained for neutrinos, $(14:11:11)_\oplus$, and antineutrinos, $(4:7:7)_\oplus$. Interestingly, this keeps the flavor ratio of the neutrino plus antineutrino flux equal to the canonical one. Departures from these canonical flavor ratios could imply non-standard production mechanisms~\cite{Enqvist:1998un, Barenboim:2003jm, Beacom:2003nh, Hung:2003jb} and/or anomalous propagation effects~\cite{Athar:2000yw, Crocker:2001zs, Beacom:2002vi, Barenboim:2003jm, Beacom:2003eu, Illana:2004qc, Illana:2005pu, Anchordoqui:2005gj, Esmaili:2009fk, Bhattacharya:2009tx, Bhattacharya:2010xj,Baerwald:2012kc,Pakvasa:2012db}, but there could also be related to uncertainties in the detection process, as unaccounted backgrounds, systematics or misidentification of muons in the detector. 

A crucial limiting factor on our present understanding of these high-energy neutrinos is the small number of observed events. In this work we focus on the 4-year IceCube high-energy starting event (HESE, also known as contained-vertex events) data~\cite{Aartsen:2015zva}, currently 53 events (plus one event whose energy and direction cannot be unambiguously assigned). We follow and update our previous works~\cite{Mena:2014sja, Palomares-Ruiz:2014zra, Palomares-Ruiz:2015mka, Vincent:2015woa}, including the topological, spectral and angular information. Furthermore, following IceCube results, we have refined the background estimation concerning both atmospheric downgoing neutrinos, accounting for a precise calculation of their veto, and atmospheric muons, with a more accurate modeling of their spectrum. In addition to performing fits to an isotropic unbroken power-law spectrum, we also investigate other scenarios. We consider an isotropic power-law spectrum but allow for a neutrino-antineutrino asymmetry in the fluxes and in the flavor compositions (see also Refs.~\cite{Barger:2014iua, Palladino:2015uoa, Nunokawa:2016pop}), which could point to a photohadronic origin, as discussed above. We also explore mixed population scenarios, as different sources could contribute to the neutrino flux and hence, could give rise to more complicated spectral features and anisotropies in the angular distribution. Therefore, as a first step in this direction, we also study an isotropic model with two unbroken power-law spectra (see also Ref.~\cite{Chen:2014gxa}), which we compare to the single-component case, and a model such that the sources from the northern and southern hemispheres produce high-energy neutrinos characterized by different energy distributions~\cite{Aartsen:2015knd}. Because a galactic component would mostly contribute to the southern hemisphere sample, this North-South model represents a simplified galactic plus extragalactic model with different spectral distributions (see also Refs.~\cite{Neronov:2015osa, Neronov:2016bnp, Palladino:2016zoe}). 

In this work, we do not include the recently-presented through-going muon sample~\cite{Aartsen:2015rwa}. Given the current publicly available data, these events cannot be modeled, outside the IceCube collaboration, in the same level of detail as the HESE events (e.g., the deposited energy of the muons is given in terms of a proxy, which is not precisely described in terms of a physical quantity), although such an analysis would be desirable. A HESE-only analysis is nonetheless crucial. Different data sets have different backgrounds, different types of signal and different systematics, so by studying them separately, one can understand their compatibility and be guided towards possible trends in the data, as it seems it happens. Among other conclusions, our HESE-only analysis shows that the recently hinted-at North-South asymmetry~\cite{Aartsen:2015knd} is driven by the through-going muons, rather than by the HESE events, for which we find no such asymmetry.

The structure of the paper is as follows. Section~\ref{sec:analysis} contains a description of the 4-year Icecube  HESE data, together with the analysis details, such as the improvements in the background calculations, the updated energy resolution and the muon track misclassification, whose strategy is also presented here. In Sec.~\ref{sec:1pownunubar} we analyze the typical scenario with an astrophysical neutrino flux characterized by an isotropic unbroken power-law spectrum.  First, we study an isotropic neutrino flux assuming identical neutrino and antineutrino spectra. Then, we allow them to have different normalizations and flavor compositions, but a common spectral shape. Lacking any statistically significant evidence of clustering or correlations, the assumption of isotropy of the astrophysical neutrino flux is justified~\cite{Aartsen:2015zva}. To conclude Sec.~\ref{sec:1pownunubar}, we perform a fit to the neutrino and antineutrino flavor ratios separately to allow for the possibility of an asymmetry between these two populations. Sec.~\ref{sec:2powNS} presents the two-component analyses, restricting ourselves first to the simplest scenario, described by two isotropic neutrino fluxes with different power-law spectra. Such a description is similar to other existing approaches in the literature, but here we allow for freedom in the neutrino flux composition: if the sources possess a mixed population profile, one could naturally expect a difference in the neutrino flavor ratios. We finally allow for different North-South hemisphere populations, characterizing each of them by a single power-law and (free) neutrino flavor ratios. We compare our results with previous findings and summarize them in Sec.~\ref{sec:summary}.

\section{4-year Icecube HESE data and Analysis}
\label{sec:analysis}

The analyses performed in this work make use of the 4-year IceCube HESE data~\cite{Aartsen:2013bka, Aartsen:2013jdh, Aartsen:2014gkd, Aartsen:2015zva} and are based on the calculations described in Ref.~\cite{Palomares-Ruiz:2015mka}, with some improvements and updates, which are detailed in the subsections below. For a complete description of our approach, we refer the reader to Ref.~\cite{Palomares-Ruiz:2015mka}. This includes modeling of the deposited EM-equivalent energy as a function of the true neutrino energy, modeling of the effective target mass, and explicit computation of the average energy deposition rates of electromagnetic and hadronic showers, as well as of muon tracks, as a function of their energy. The calculation of neutrino propagation, attenuation and regeneration in the Earth is also described, for both astrophysical and atmospheric neutrino fluxes.  In this work, in addition to the usual isotropic single power-law spectrum, we also study different models for the astrophysical neutrino flux. Below, we define the likelihoods for these models and the set of free parameters in our analyses. The results are then presented in the following two sections and summarized in Tabs.~\ref{tab:1pow}$-$\ref{tab:NS}.

\subsection{4-year IceCube HESE data}
\label{sec:data}

The HESE data correspond to neutrino-like events whose first light is seen inside the fiducial volume of the IceCube detector. By excluding events that also trigger the surrounding veto region, this technique is sensitive to all sky and to all neutrino flavors with interaction vertices in the detector. After 1347~days of data taking, 54 events (17 in the fourth year alone) with electromagnetic (EM)-equivalent deposited energies in the range $\sim [20~{\rm TeV}-2~{\rm PeV}]$ have been detected by IceCube. One of these events (already present in the 3-year sample) is a pair of coincident muon tracks whose energy and direction cannot be unambiguously assigned and is therefore not included in our analysis.

At these energies, there are two possible event topologies in the IceCube detector\footnote{We do not consider here double-bang events~\cite{Learned:1994wg}. The expected number of events of this type in IceCube after 4~years, for neutrino fluxes compatible with the observed spectrum, is less than one~\cite{Aartsen:2015dlt, Palladino:2015uoa}.}: muon tracks and showers (electromagnetic or hadronic or both). Among the 53 events in the entire sample, 14 are muon tracks and 39 are showers. 16 of the events come from the northern hemisphere (5 tracks) and 37 from the southern hemisphere (9 tracks). We also consider another EM-equivalent deposited energy interval with a higher energy threshold, [60~TeV $-$ 10~PeV], yielding a smaller but cleaner (low-background) sample. In this case, there are  32 events (8 tracks and 24 showers): 10 events come from the northern hemisphere (4 tracks) and 22 from the southern hemisphere (4 tracks).

\subsection{Improvements on the calculation of the backgrounds}
\label{sec:bkg}

The two main sources of background for HESE are atmospheric muons and neutrinos, produced as secondaries in cosmic-ray interactions with the nuclei of the Earth's atmosphere. The expected number of atmospheric muon and neutrino events in the entire energy range, after 4 years of data, is $N_\mu = 12.6 \pm 5.1$ and $N_\nu = 9.0^{+8.0}_{-2.2}$~\cite{Aartsen:2015zva}, respectively. Here, we describe the improvements on the calculation of these backgrounds with respect to our previous work~\cite{Palomares-Ruiz:2015mka}. 

As in Ref.~\cite{Palomares-Ruiz:2015mka}, we consider the calculation of the conventional (mainly from $\pi$ and $K$ decays) and prompt (from charmed meson decays) atmospheric neutrino fluxes of Refs.~\cite{Sinegovsky:2011ab, Petrova:2012qf, Sinegovskaya:2013wgm, Sinegovskaya:2014pia} based on the Hillas and Gaisser (HGm) cosmic-ray approximation~\cite{Hillas:2006ms, Gaisser:2012zz} and the hadronic model of Kimel and Mokhov (KM)~\cite{Kalinovsky:1989kk}, with updated parameters~\cite{Naumov:2001uc, Fiorentini:2001wa, Naumov:2002dm} for the conventional flux. We use the Zatsepin and Sokolskaya cosmic-ray model (ZS)~\cite{Zatsepin:2006ci} and the quark-gluon string model (QGSM)~\cite{Kaidalov:1984ne, Kaidalov:1986zs, Kaidalov:1985jg, Bugaev:1989we} for the prompt flux. These fluxes are similar, for the energies of interest, to the ones used by the IceCube collaboration~\cite{Honda:2006qj, Enberg:2008te, Bhattacharya:2015jpa}.

One of the improvements we have implemented is in the calculation of the veto for downgoing atmospheric neutrinos. The suppression of this background flux is possible by tagging muons produced in the same cosmic-ray cascade, which can trigger the muon veto of the detector~\cite{Schonert:2008is, Gaisser:2014bja}. In this work, we have used the code provided as Supplemental Material of Ref.~\cite{Gaisser:2014bja}, which computes the veto probability for conventional and prompt atmospheric neutrinos using both, the correlated muons produced by the same parent meson decay as the muon neutrino~\cite{Schonert:2008is} and the uncorrelated muons from other branches of the cascade~\cite{Gaisser:2014bja}. Whereas the correlated veto only applies to muon neutrinos, the uncorrelated veto (not incorporated in Ref.~\cite{Palomares-Ruiz:2015mka}) applies to electron neutrinos too. In addition, we have updated the minimum muon energy required when reaching the detector. Instead of $E_{\rm th} = 10$~TeV~\cite{Aartsen:2013jdh}, we use $E_{\rm th} = 1$~TeV, which represents a much better description of the threshold~\cite{Gaisser:2014bja, Kopper}. Finally, unlike what was done in the first IceCube analyses~\cite{Aartsen:2013jdh} and in Ref.~\cite{Palomares-Ruiz:2015mka}, we have not limited the maximum possible suppression to 90\%~\cite{Kopper}.

Another improvement is the modeling of the atmospheric muon background,  which allows us to use the whole energy range, i.e., including events with EM-equivalent deposited energies as low as 10~TeV. We use the spectral information contained in the plots presented by the IceCube collaboration for the 3-year data~\cite{Aartsen:2014gkd} and fitted the following function:
\begin{equation}
F_\mu(E) = \int_E^\infty \frac{dN_\mu (E_{\rm dep})}{dE_{\rm dep}} \, dE_{\rm dep} = 
A_{3yr} \, \frac{1 + p \, (E/E_1)^g}{1 + q \, (E/E_1)^{h_0 + h_1 (E/E_2)^c}} ~,
\label{eq:muonfit}
\end{equation}
where $E$ is expressed in TeV, the normalization constant $A_{3yr}$ is chosen so that $F_\mu(10~{\rm TeV})=8.4$ (the predicted number of atmospheric muons in the 3-year sample), $E_1 = 30$~TeV, $E_2 = 60$~TeV, and the fitting constants are: $c=6.0$, $p=1.3$, $q=5.1$, $g=24.8$, $h_0=27.0$ and $h_1=0.10$. Thus, for a differential spectrum that asymptotes to zero for very large energies, as the atmospheric muon background,
\begin{equation}
\frac{dN_\mu (E_{\rm dep})}{dE_{\rm dep}} = - \frac{dF_\mu (E_{\rm dep})}{dE_{\rm dep}} ~.
\label{eq:muonfit2}
\end{equation}
This flux is correspondingly scaled, so that the predicted number of atmospheric muons after 4 years is $N_\mu=12.6$~\cite{Aartsen:2015zva}.

\subsection{Energy resolution and track misidentification}
\label{sec:res}

Using the extra information from the events of the fourth year, we have also updated the fit to the energy resolution function, represented by a Gaussian of width $\sigma(E_{\rm true})$. As done in Ref.~\cite{Palomares-Ruiz:2015mka}, we assume the uncertainty on the true EM-equivalent deposited energy, $E_{\textrm{true}}$, to be given by the error on the measured EM-equivalent deposited energy, $E_{\textrm{dep}}$. Here we perform a one-parameter fit, within the observed energy range, using the 53 (shower and track) events detected by IceCube after 1347~days, and we obtain a dispersion $\sigma(E_{\textrm{true}}) = 0.118 \ E_{\textrm{true}}$.

Moreover, throughout  this work, we assume that 30\% of the neutrino-induced muon tracks in the detector are misclassified as showers, either because  the muon deposits too little energy in the detector or because it is produced near the edges and escapes detection~\cite{Aartsen:2015ivb}. Although atmospheric muons are usually detected as tracks, they could also undergo catastrophic energy losses and be misidentified as showers. As done in Ref.~\cite{Palomares-Ruiz:2015mka}, we set this misclassification fraction to 10\%~\cite{Aartsen:2014gkd}. These are oversimplifications, as the fraction of misclassified tracks depends on the muon energy~\cite{Aartsen:2014muf}. However, more detailed information is not publicly available.

\subsection{Statistical analysis}
\label{sec:likelihood}

We perform unbinned extended maximum likelihood analyses, using the energy,  event topology and  hemisphere of origin information of all the 53 high-energy starting events detected after 4 years of data taking in IceCube. Dropping an overall multiplicative factor, the full likelihood is given by:
\begin{equation}
\like = e^{- \sum_f N_f}  \, \prod_{i = 1}^{N_{\rm obs}} \like_i ~,
\end{equation}
where $N_{\rm obs} = N^{\rm sh}_{\rm N} + N^{\rm sh}_{\rm S} + N^{\rm tr}_{\rm N} + N^{\rm tr}_{\rm S}$ is the total number of observed events, $N_f$ refers to the total number of predicted events from source $f$ and the sum runs over the four type of sources we consider: $f = $ \{astrophysical neutrinos ($a$),  atmospheric muon events ($\mu$), conventional atmospheric neutrinos ($\nu$) and prompt atmospheric neutrinos ($p$)\}. However, note that, depending on the analysis we perform, we might split the contribution of some of the sources into more than one term. We define $N^{\textrm{sh}}_{\rm N}$, $N^{\textrm{sh}}_{\rm S}$, $N^{\textrm{tr}}_{\rm N}$ and $N^{\textrm{tr}}_{\rm S}$ as the observed number of upgoing (from the North hemisphere) showers, downgoing (from the South hemisphere) showers, upgoing tracks and downgoing tracks, respectively (their sum is equal to $N_{\rm obs}$). Each observed event $i$ is identified by the set $\{E_{\rm dep,i}, H_i, T_i\}$, which indicate: $E_{\mathrm{dep},i}$: ~EM-equivalent deposited energy; $H_i $: direction of origin (\{North, South\}); and $T_i$:~topology (\{track, shower\}). The likelihood for each event is the sum of the likelihoods for each of the possible sources and is given by
\begin{equation}
\like_i = \sum_f N_f \, \pdf_f(\{E_{\mathrm{dep},i},H_i,T_i\})~,
\label{eq:fullLike}
\end{equation}
where similarly to Ref.~\cite{Palomares-Ruiz:2015mka}, we define the normalized probability density function (PDF) for each event $i$ and for each source $f$, $\pdf_f(\{E_{\mathrm{dep},i},H_i,T_i\})$, as the probability distribution for an observed event with EM-equivalent deposited energy $E_{\textrm{dep}, i}$ and topology $T_i$ originated from a flux of type $f$ of incoming neutrinos from hemisphere $H_i$. 

Throughout this work, we define (each component of) the all-flavor astrophysical neutrino and antineutrino flux as
\begin{equation}
\frac{d\Phi^a}{dE_\nu} = \phi_H \, \left(\frac{E_\nu}{100 \, {\rm TeV}}\right)^{-\gamma_H}
\end{equation}
where $\gamma_H$ is the hemisphere-dependent spectral index and $\phi_H$ is the hemisphere-dependent flux normalization, which is given in units of $10^{-18} \, {\rm GeV}^{-1} \, {\rm cm}^{-2} \, {\rm s}^{-1} \, {\rm sr}^{-1}$. 

Therefore, the PDF of a given event $i$ for a flux of astrophysical neutrinos coming from hemisphere $H_i$ to be produced with EM-equivalent deposited energy $E_{\textrm{dep}, i}$, given the flavor combination at Earth $\{\alpha_{e,H_i} : \alpha_{\mu,H_i} : \alpha_{\tau,H_i}\}_\oplus$ and spectrum $d\Phi^a/dE_\nu$, can be written as
\begin{equation}
\pdf_f(\{E_{\mathrm{dep},i},H_i,T_i\}; \{\alpha\}_\oplus,\phi,\gamma)= \frac{1}{\sum_{\ell, H', T'} (\alpha_{\ell,H'})_\oplus \int_\Emin^\Emax dE_{\textrm{dep}} \, \frac{d\left(N_a\right)_{\ell, H'}^{T'}}{dE_{\textrm{dep}} }} \, \sum_\ell (\alpha_{\ell,H})_\oplus \, \frac{d\left(N_{a}\right)_{\ell, H_i}^{T_i}}{dE_{\textrm{dep},i}}   ~.
\label{eq:PDFa}
\end{equation}
where $\Emin$ and $\Emax$ are the minimum and maximum EM-equivalent deposited energies in each analysis, and the sum in the denominator goes over the three neutrino flavors, $\ell = \{e, \mu, \tau\}$, the hemisphere, $H' =$\{N, S\}, and the type of event topology, $T'=\{\textrm{tr, sh}\}$. The event spectrum $d\left(N_a\right)_{\ell, H}^{T}/dE_{\textrm{dep}}$ results from the sum of all the partial contributions from different processes to topology $T$ from neutrinos and antineutrinos (except in the case when the neutrino and antineutrino fluxes are fitted separately) of flavor $\ell$ from direction $H$ (for more details, see Ref.~\cite{Palomares-Ruiz:2015mka}).

Likewise, for the atmospheric backgrounds we have
\begin{eqnarray}
\pdf_\nu(\{E_{\mathrm{dep},i},H_i,T_i\}; N_\nu) & = & \frac{1}{\sum_{H', T'} \int_\Emin^\Emax dE_{\textrm{dep}} \,  \frac{d\left(N_\nu\right)_{H'}^{T'}}{dE_{\textrm{dep}} }} \, \frac{d\left(N_\nu\right)_{H_i}^{T_i}}{dE_{\textrm{dep},i}}  \hspace{1cm} \textrm{(conventional atmospheric $\nu$)} ~, 
\label{eq:PDFnu} \\
\pdf_\mu(\{E_{\mathrm{dep},i},H_i,T_i\}; N_\mu) & = & \frac{1}{\sum_{H',T'} \int_\Emin^\Emax dE_{\textrm{dep}} \,  \frac{d\left(N_\mu\right)_{H'}^{T'}}{dE_{\textrm{dep}} }} \, \frac{d\left(N_\mu\right)_{H_i}^{T_i}}{dE_{\textrm{dep},i}}  \hspace{1cm} \textrm{(atmospheric $\mu$)} ~, \label{eq:PDFmu} \\
\pdf_p(\{E_{\mathrm{dep},i},H_i,T_i\}; N_p) & = & \frac{1}{\sum_{H', T'} \int_\Emin^\Emax dE_{\textrm{dep}} \,  \frac{d\left(N_p\right)_{H'}^{T'}}{dE_{\textrm{dep}} }} \, \frac{d\left(N_p\right)_{H_i}^{T_i}}{dE_{\textrm{dep},i}} \hspace{1cm} \textrm{(prompt atmospheric $\nu$)}  ~,
\label{eq:PDFp}
\end{eqnarray}
where the conventional and prompt atmospheric neutrino fluxes are used to compute the event distributions, analogously to the astrophysical case. For the conventional and prompt atmospheric neutrino fluxes, the relative flavor and $\nu/\bar \nu$ contributions are fixed to the predicted theoretical flux for each case. For these background sources,  the sum over flavors of the differential event spectra includes neutrinos and antineutrinos. The event distribution of atmospheric muons is modeled as discussed above. 

Throughout this work, and following IceCube analyses~\cite{Aartsen:2014gkd, Aartsen:2015zva}, we let all fitted parameters vary freely, but we impose a Gaussian prior on the atmospheric muon background, since the expected number of veto-passing muons comes from an independent measurement~\cite{Aartsen:2013jdh, Aartsen:2014gkd}. For each of the two EM-equivalent deposited energy intervals we consider, the central values and the 1$\sigma$ C.L. uncertainties we use are:  $N_\mu = 12.6 \pm 5.1$ for [10~TeV $-$ 10~PeV] and $N_\mu = 0.6 \pm 0.3$ for [60~TeV $-$ 10~PeV]. We perform different type of analyses using the \textsc{MultiNest} nested sampling algorithm~\cite{Feroz:2007kg, Feroz:2008xx, Feroz:2013hea}. Where possible, we present our results as profile likelihoods (orange and purple), and use the log-likelihood ratio as the test statistic assuming it has a $\chi^2$ distribution, to facilitate comparison with the IceCube results. However, when the number of fitted parameters becomes too large to sample efficiently, we instead present bayesian posterior distributions (blue and green). Where possible, we have verified that our 68\% and 95\% credible regions are in excellent agreement with the corresponding 1$\sigma$ C.L. and 2$\sigma$ C.L. profile likelihood contours.

\begin{sidewaystable}
	\caption{\textbf{\textit{Isotropic single power-law model: Best-fit points (Bayesian posterior means in parentheses)}}, with $1\sigma$ errors (posterior means with 68\% lower and upper containment levels). Fixed quantities are indicated by italics. In the 5P fit, we set $\alpha_{\mu,\oplus} = \alpha_{\tau, \oplus} = (1 - \alpha_{e, \oplus})/2$.}
	\label{tab:1pow}
	\begin{tabular*}{\textwidth}{c @{\extracolsep{\fill}} c c   c   c   c   c  c }
		\hline
		Energy range & Model &  $(\alpha_e:\alpha_\mu:\alpha_\tau)_{\oplus}$ &$\gamma$ & $\phi$ & $N_\mu$ & $N_\nu$  & $N_p$  \\ \hline
		\multirow{4}{*}{10~TeV$-$10~PeV} & 4P & \textit{(1 : 1 : 1)} &  2.87$^{+0.23}_{-0.26}$ (2.86$_{-0.18}^{+0.18}$) & 8.6$^{+2.4}_{-2.4}$ (8.2$_{-1.8}^{+1.9}$) & 9.5$^{+3.6}_{-3.0}$ (9.7$_{-2.5}^{+2.5}$) & 8.7$^{+7.9}_{-5.7}$ (11.6$_{-5.2}^{+5.1}$)  & \textit{0}  \\
		& 5P & $(0.06 : 0.47 : 0.47)$ &  2.97$^{+0.25}_{-0.27}$ (2.86$_{-0.20}^{+0.19}$) & 14.4$^{+4.8}_{-6.3}$ (9.3$_{-3.2}^{+3.2}$) & 9.1$^{+3.5}_{-3.1}$ (9.8$_{-2.6}^{+2.6}$) & 2.1$^{+10.1}_{-2.1}$ (11.3$_{-6.1}^{+6.1}$)  & \textit{0}  \\
		& 6P & $(0.03 : 0.37 : 0.60)$  & 2.84$^{+0.25}_{-0.27}$ (2.86$_{-0.18}^{+ 0.18}$) & 11.1$^{+3.7}_{-4.8}$ (8.9$_{-2.6}^{+ 2.4}$) & 9.2$^{+3.7}_{-3.0}$ (9.8$_{-2.6}^{+2.6}$) & 8.7$^{+4.9}_{-2.1}$ (11.7$_{-3.4}^{+ 3.5}$) & \textit{0}  \\
		& 7P & $(0.05 : 0.41 : 0.54)$ & 2.95$^{+0.26}_{-0.30}$ (2.76$_{-0.26}^{+0.25}$) & 13.2$^{+4.6}_{-5.6}$ (7.0$_{-3.2}^{+3.2}$) & 9.2$^{+3.2}_{-3.4}$ (9.9$_{-2.6}^{+ 2.5}$) & 3.4$^{+8.5}_{-3.4}$ (10.1$_{-5.3}^{+ 5.5}$) & 0.3$^{+6.0}_{-0.3}$ (9.4$_{-6.1}^{+6.4}$)  \\ \hline
		\multirow{4}{*}{60~TeV$-$10~PeV} & 4P & \textit{(1 : 1 : 1)} & 2.72$^{+0.30}_{-0.29}$ (2.73$_{-0.21}^{+0.22}$) & 6.5$^{+3.2}_{-2.5}$ (6.6$_{-2.3}^{+2.3}$) & 0.7$^{+0.5}_{-0.7}$ (0.7$_{-0.2}^{+0.2}$) & 7.3$^{+5.9}_{-5.0}$ (9.8$_{-4.5}^{+4.5}$) & \textit{0} \\
		& 5P & $(0.10 : 0.45 : 0.45)$ &  2.75$^{+0.36}_{-0.26}$ (2.72$_{-0.23}^{+0.23}$) & 9.3$^{+7.0}_{-5.6}$ (6.9$_{-2.9}^{+3.2}$) & 0.7$^{+0.5}_{-0.7}$ (0.7$_{-0.2}^{+0.2}$) & 4.8$^{+7.7}_{-4.7}$ (9.7$_{-4.8}^{+4.8}$)  & \textit{0}  \\
		& 6P & $(0.00 : 0.40 : 0.60)$ & 2.77$^{+0.31}_{-0.42}$ (2.72$_{-0.23}^{+0.24}$) & 10.3$^{+7.1}_{-6.3}$ (7.0$_{-2.9}^{+2.9}$) & 0.7$^{+0.5}_{-0.7}$ (0.7$_{-0.2}^{+0.2}$) & 4.3$^{+8.2}_{-4.3}$ (9.7$_{-4.4}^{+4.4}$)  & \textit{0}  \\
		& 7P & $(0.18 : 0.52 : 0.30)$ &  2.75$^{+0.36}_{-0.34}$ (2.55$_{-0.31}^{+0.31}$) & 9.0$^{+6.9}_{-5.1}$ (4.8$_{-2.8}^{+ 2.8}$) & 0.7$^{+0.5}_{-0.7}$ (0.7$_{-0.2}^{+0.2}$) & 5.3$^{+7.0}_{-5.3}$ (8.4$_{-4.3}^{+ 4.2}$) & 0.1$^{+5.0}_{-0.1}$ (7.5$_{-4.6}^{+4.6}$) \\ \hline
	\end{tabular*}
	
	\caption{\textbf{\textit{Neutrino-antineutrino model: Best-fit points (Bayesian posterior means in parentheses)}}, with $1\sigma$ errors (posterior means with 68\% lower and upper containment levels).  In this case, 7P+$\nu \bar \nu$, we assume the same spectral index for neutrinos and antineutrinos, $\gamma=\bar\gamma$, but allow for a different neutrino-antineutrino flavor composition, with $\alpha_{\mu, \oplus} = \alpha_{\tau, \oplus} = (1 - \alpha_{e, \oplus})/2$ and $\bar\alpha_{\mu, \oplus} = \bar\alpha_{\tau, \oplus} = (1 - \bar\alpha_{e, \oplus})/2$.}
	\label{tab:nunubar}
	\begin{tabular*}{\textwidth}{c @{\extracolsep{\fill}} c c c c c c c c }
		\hline
		Energy range & Model & $\alpha_{e, \oplus}$ & $\bar \alpha_{e, \oplus}$ & $\gamma$ & $\phi$ & $\bar \phi$ & $N_\mu$ & $N_\nu$  \\ \hline
		10~TeV$-$10~PeV & \multirow{2}{*}{7P+$\nu\bar\nu$} & 0.08 &  0.39 & 2.91$^{+0.31}_{-0.24}$ (2.87$_{-0.20}^{+0.19}$) & 10.0$^{+2.8}_{-10.0}$ (4.8$_{-2.5}^{+ 2.6}$) & 0.3$^{+11.8}_{-0.3}$ (3.9$_{-2.3}^{+ 2.5}$) & 9.3$^{+3.1}_{-3.5}$ (9.8$_{-2.5}^{+2.6}$) & 4.2$^{+8.2}_{-4.2}$ (10.8$_{-5.6}^{+5.6}$)   \\
		60~TeV$-$10~PeV & & 0.12 & 1.00 &  2.74$^{+0.33}_{-0.35}$ (2.74$_{-0.23}^{+0.21}$) & 6.5$^{+5.5}_{-6.5}$ (3.7$_{-1.9}^{+2.0}$) & 0.1$^{+10.6}_{-0.1}$ (3.2$_{-2.0}^{+1.9}$) & 0.7$^{+0.5}_{-0.7}$ (0.7$_{-0.2}^{+0.2}$) & 5.2$^{+7.6}_{-5.2}$ (9.1$_{-4.2}^{+4.2}$)  \\ \hline
	\end{tabular*}
	
	\caption{\textbf{\textit{Isotropic two power-law model: Best-fit points (Bayesian posterior means in parentheses)}}, with $1\sigma$ errors (posterior means with 68\% lower and upper containment levels). In this case, 8P+2pow, we impose $(\alpha_{\mu,\rm s})_\oplus = (\alpha_{\tau,\rm s})_\oplus = (1 - (\alpha_{e,\rm s})_\oplus)/2$ and $(\alpha_{\mu,\rm h})_\oplus = (\alpha_{\tau,\rm h})_\oplus = (1 - (\alpha_{e,\rm h})_\oplus)/2$.}
	\label{tab:2pow}
	\begin{tabular*}{\textwidth}{c @{\extracolsep{\fill}} c c c c c c c }
		\hline
		Energy range & Model & & $(\alpha_{e, \rm s})_\oplus/(\alpha_{e, \rm h})_\oplus$ & $\gamma_{\rm s}/\gamma_{\rm h}$ & $\phi_{\rm s}/\phi_{\rm h}$ & $N_\mu$ & $N_\nu$  \\ \hline
		\multirow{2}{*}{10~TeV$-$10~PeV} & \multirow{4}{*}{8P+2pow} & soft & 0.08 &  3.50$^{+1.55}_{-0.41}$ (3.69$_{-0.52}^{+0.55}$) & 12.9$^{+4.9}_{-12.9}$ (4.7$_{-2.8}^{+3.1}$) &  \multirow{2}{*}{9.4$^{+2.7}_{-4.0}$ (9.4$_{-2.4}^{+ 2.4}$)} & \multirow{2}{*}{1.6$^{+9.4}_{-1.6}$ (10.0$_{-5.3}^{+ 5.4}$)}  \\
		& & hard & 0.00 & 2.09$^{+0.92}_{-0.64}$ (2.44$_{-0.34}^{+0.34}$) & 1.4$^{+12.7}_{-1.4}$ (4.0$_{-2.8}^{+2.9}$) & & \\
		\multirow{2}{*}{60~TeV$-$10~PeV} & & soft & 0.01 & 3.89$^{+1.08}_{-1.16}$ (3.65$_{-0.61}^{+0.60}$) & 13.8$^{+8.1}_{-13.8}$ (5.8$_{-3.3}^{+3.5}$) & \multirow{2}{*}{0.7$^{+0.5}_{-0.7}$ (0.7$_{-0.2}^{+0.2}$)} & \multirow{2}{*}{1.8$^{+9.1}_{-1.8}$ (8.6$_{-4.1}^{+4.2}$)}  \\ 
		& & hard & 0.03 & 2.25$^{+0.58}_{-0.81}$ (2.28$_{-0.34}^{+0.35}$) & 2.7$^{+8.5}_{-2.7}$ (2.6$_{-1.8}^{+2.0}$) &  &  \\ \hline 
	\end{tabular*}
	
	\caption{\textbf{\textit{North-South model: Best-fit points (Bayesian posterior means in parentheses)}}, with $1\sigma$ errors (posterior means with 68\% lower and upper containment levels). In this case, 8P+NS, we set $(\alpha_{\mu,\rm N})_\oplus = (\alpha_{\tau,\rm N})_\oplus = (1 - (\alpha_{e,\rm N})_\oplus)/2$ and $(\alpha_{\mu,\rm S})_\oplus = (\alpha_{\tau,\rm S})_\oplus = (1 - (\alpha_{e,\rm S})_\oplus)/2$.}
	\label{tab:NS}
	\begin{tabular*}{\textwidth}{c @{\extracolsep{\fill}} c c c c c c c }
		\hline
		Energy range & Model & & $(\alpha_{e,\rm N})_\oplus/(\alpha_{e,\rm S})_\oplus$ & $\gamma_{\rm N}/\gamma_{\rm S}$ & $\phi_{\rm N}/\phi_{\rm S}$ &  $N_\mu$ & $N_\nu$ \\ \hline
		\multirow{2}{*}{10~TeV$-$10~PeV} & \multirow{4}{*}{8P+NS} & North & 0.00 & 2.96$^{+0.41}_{-0.78}$ (2.83$_{-0.51}^{+0.49}$) & 13.3$^{+3.3}_{-11.1}$ (4.8$_{-2.9}^{+3.0}$) & \multirow{2}{*}{9.1$^{+3.7}_{-3.1}$ (9.2$_{-2.5}^{+2.6}$)} & \multirow{2}{*}{4.1$^{+13.5}_{-4.1}$ (15.2$_{-5.7}^{+ 6.0}$)}  \\
		& & South & 0.26 & 2.94$^{+0.24}_{-0.29}$ (2.90$_{-0.19}^{+ 0.20}$) & 12.0$^{+5.6}_{-5.5}$ (9.5$_{-3.0}^{+3.0}$) &  & \\
		\multirow{2}{*}{60~TeV$-$10~PeV} & & North & 0.04 & 2.42$^{+0.89}_{-0.61}$ (2.73 $_{-0.65}^{+0.67}$) & 3.7$^{+10.2}_{-3.7}$ (3.9$_{-2.6}^{+2.7}$) & \multirow{2}{*}{0.6$^{+0.5}_{-0.6}$ (0.7$_{-0.2}^{+0.2}$)} & \multirow{2}{*}{8.1$^{+6.4}_{-6.5}$ (11.1$_{-4.1}^{+3.8}$)}  \\
		& & South & 0.28 & 2.79$^{+0.31}_{-0.29}$ (2.81$_{-0.23}^{+0.24}$) & 8.7$^{+6.3}_{-4.7}$ (7.9$_{-2.6}^{+3.0}$) & & \\ \hline
	\end{tabular*}
\end{sidewaystable}

For the cases where we consider the astrophysical flux to be a single isotropic power-law component, we vary all the flavor fractions ($\alpha_{e, \oplus}$ and $\alpha_{\mu, \oplus}$, with $\alpha_{\tau, \oplus} = 1 - \alpha_{e, \oplus} - \alpha_{\mu, \oplus}$), the spectral index of the astrophysical flux ($\gamma$), and the the total number of events for all the fluxes ($N_a$, $N_\mu$, $N_\nu$ and $N_p$). In Ref.~\cite{Palomares-Ruiz:2015mka}, these fits are referred to as the ``7P'' fits, or ``6P'' fits when we set $N_p=0$. In addition, we also consider the possibility that the normalization and the flavor compositions of the astrophysical neutrino and antineutrino fluxes might not be the same. In order not to significantly increase the number of free parameters, we set $\alpha_{\mu, \oplus} = \alpha_{\tau, \oplus}$ (and equivalently for antineutrinos, $\bar \alpha_{\mu, \oplus} = \bar \alpha_{\tau, \oplus}$), which is expected from averaged neutrino oscillations, so that we have 7 free parameters $\{\alpha_{e, \oplus}, \bar \alpha_{e, \oplus}, \gamma, N_a, \bar N_a, N_\mu, N_\nu \}$, and we refer to this fit as ``7P+$\nu\bar \nu$". In this case, the event distributions in Eq.~(\ref{eq:PDFa}) correspond to neutrinos or antineutrinos, but not to their sum, so that the astrophysical PDF is split into two, one for neutrinos and one for antineutrinos.

Moreover, we consider the possibility of a two-component astrophysical flux. First, we consider the case of an isotropic two power-law astrophysical flux. Again, we set $(\alpha_{\mu, \rm s})_\oplus = (\alpha_{\tau, \rm s})_\oplus$ and $(\alpha_{\mu, \rm h})_\oplus = (\alpha_{\tau, \rm h})_\oplus$, where the indices `s' and `h' respectively refer to the soft and hard components. In this case, ``8P+2pow", the set of free parameters is $\{(\alpha_{e, \rm s})_\oplus, (\alpha_{e, \rm h})_\oplus, \gamma_{\rm s}, \gamma_{\rm h}, N_{a, \rm s}, N_{a, \rm h}, N_\mu, N_\nu \}$. Finally, we also perform a fit allowing the astrophysical fluxes from the northern and southern hemispheres to be independent of each other. In this case, there is no sum over $\lambda$ in the denominator of the PDF's defined in Eq.~(\ref{eq:PDFa}), but $H' = $~N or S. Likewise, we set $(\alpha_{\mu, \rm N})_\oplus = (\alpha_{\tau, \rm N})_\oplus$ (for the northern, and equivalently for the southern hemisphere, $(\alpha_{\mu, \rm S})_\oplus = (\alpha_{\tau, \rm S})_\oplus$), so that we have 8 free parameters $\{(\alpha_{e, \rm N})_\oplus, (\alpha_{e, \rm S})_\oplus, \gamma_{\rm N}, \gamma_{\rm S}, N_{a, \rm N}, N_{a, \rm S}, N_\mu, N_\nu \}$, and we refer to this fit as ``8P+NS".

\begin{figure}
	\begin{tabular}{l l}
		\hspace{-5mm} 
		\includegraphics[width=0.55\textwidth]{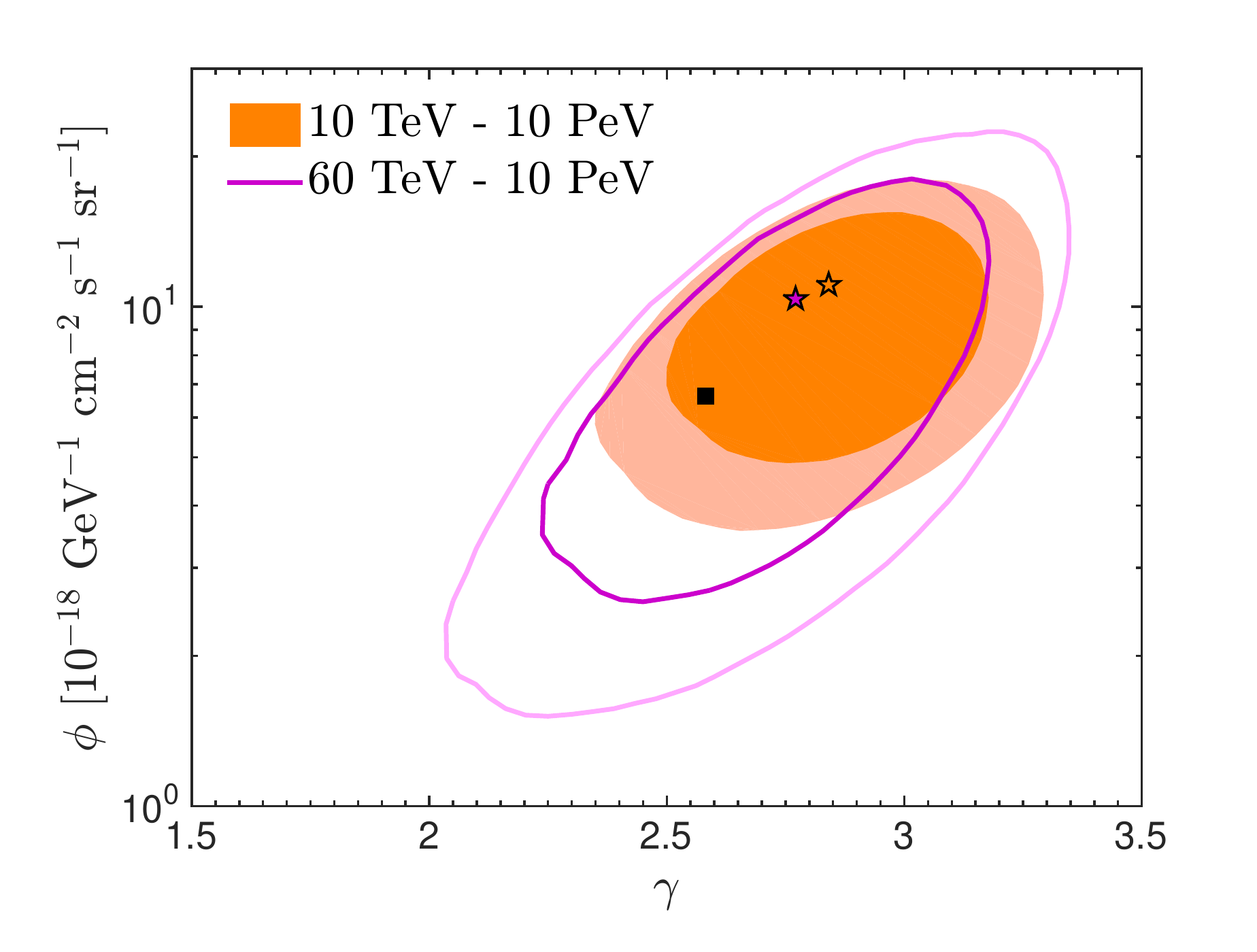} & \hspace{-10mm} \includegraphics[width=0.55\textwidth]{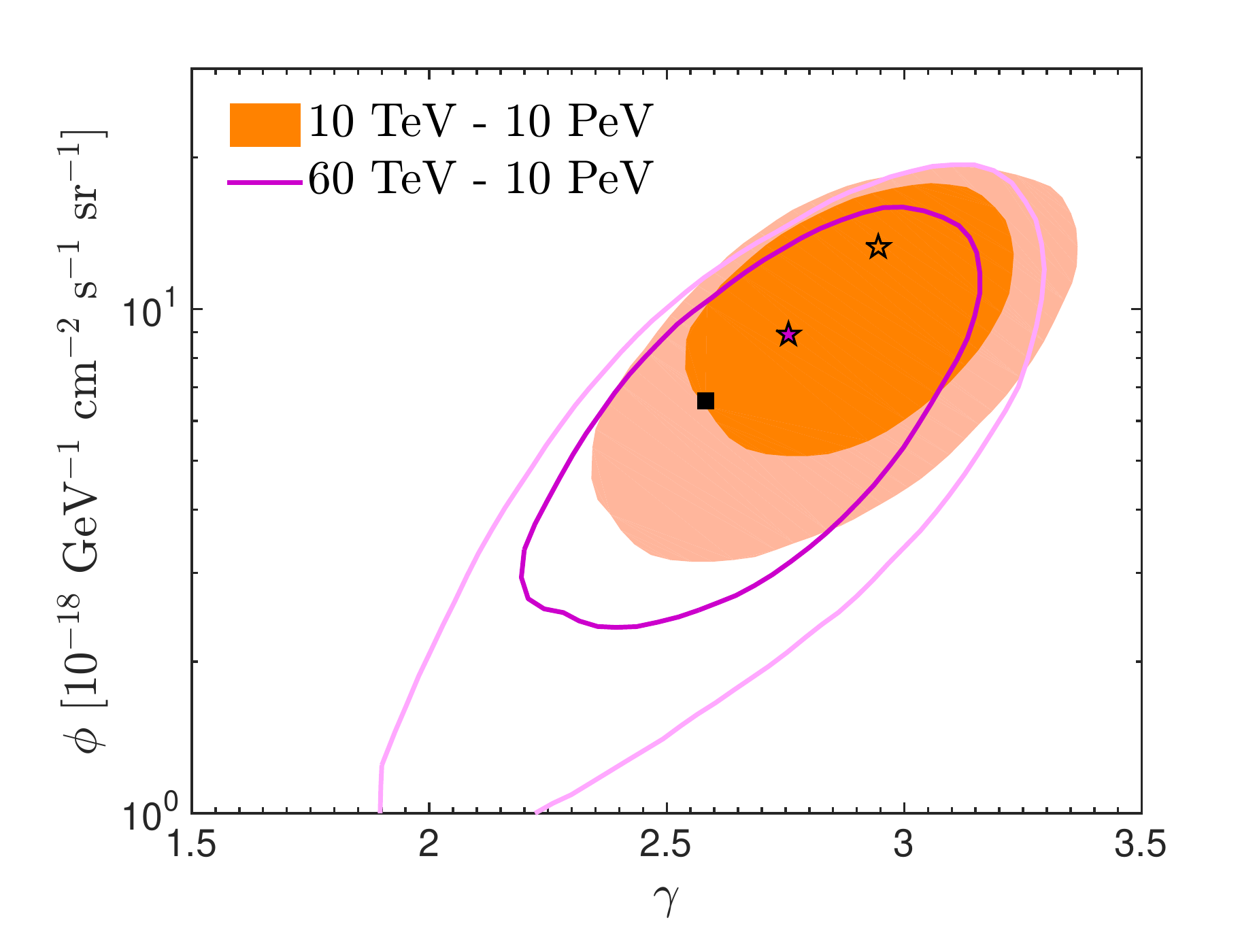} 
	\end{tabular}
	\caption{\textbf{\textit{Isotropic single power-law model: spectral shape.}} Profile likelihood contours in the astrophysical neutrino $\gamma - \phi$ plane, at $68\%$ C.L. (dark colors) and $95\%$ C.L. (light colors). Filled orange contours (closed purple curves) represent the EM-equivalent deposited energy interval of [10~TeV $-$ 10~PeV] ([60~TeV $-$ 10~PeV]). Our best fits (stars) and the preliminary IceCube best fit for $(1:1:1)_\oplus$ in the EM-equivalent deposited energy interval [60~TeV $-$ 3~PeV]~\cite{Aartsen:2015zva} (square), $E_\nu^2 \, d\Phi/dE_\nu = 6.6 \times 10^{-8} \, (E_\nu/100 \, {\rm TeV})^{-0.58}  \, {\rm GeV}^{-1} \, {\rm cm}^{-2} \, {\rm s}^{-1} \, {\rm sr}^{-1}$, are also indicated. \textit{Left panel}: without including a prompt atmospheric neutrino component (6P analysis). \textit{Right panel:} including a prompt atmospheric neutrino component (7P analysis).}
	\label{fig:norm6P}
\end{figure}

\section{Single power-law analysis}
\label{sec:1pownunubar}

In this section we consider an astrophysical neutrino flux with an isotropic unbroken power-law spectrum.  First, we study the case of identical neutrino and antineutrino fluxes. In Sec.~\ref{sec:nunubar}, we allow them to have different normalizations and flavor compositions, but with the same spectral shape. Lacking statistically significant evidence of clustering or correlations, the assumption of isotropy of the astrophysical neutrino flux is justified. An all-sky clustering test yields $p$-value = 0.58 and a test of a possible correlation with the galactic plane, allowing its width to vary, yields $p$-value = 0.025 ($p$-value=0.07 if the width is fixed to $2.5^\circ$)~\cite{Aartsen:2015zva}. The results of this section are summarized in Tabs.~\ref{tab:1pow} and~\ref{tab:nunubar}, where we quote the normalization of the fluxes ($\phi$), instead of the total number of events ($N$).

\subsection{Isotropic power-law model}
\label{sec:1pow}

\begin{figure}
	\begin{tabular}{l l}
		\hspace{-12mm}
		\includegraphics[width=0.55\textwidth]{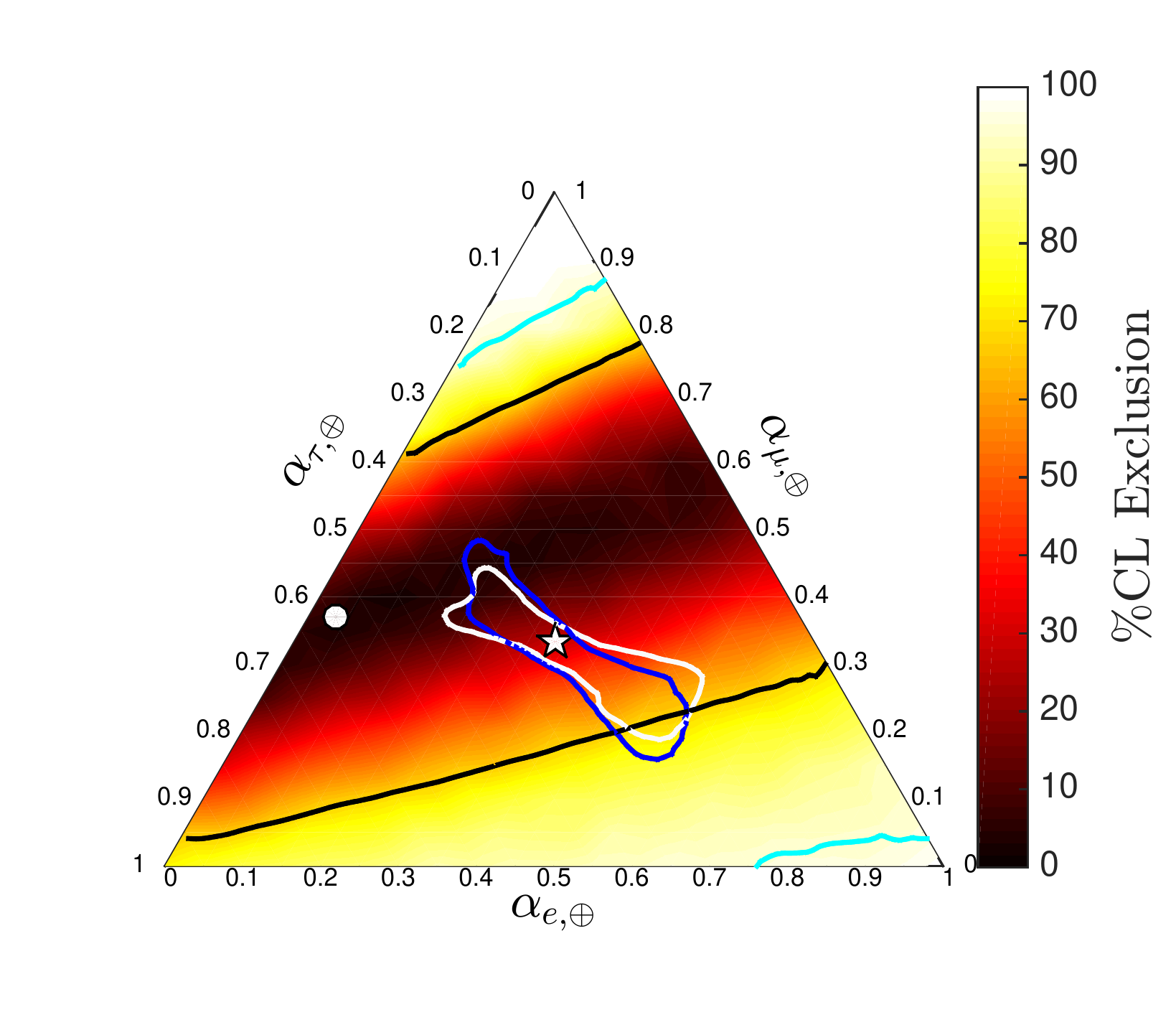} & \hspace{-7mm} \includegraphics[width=0.55\textwidth]{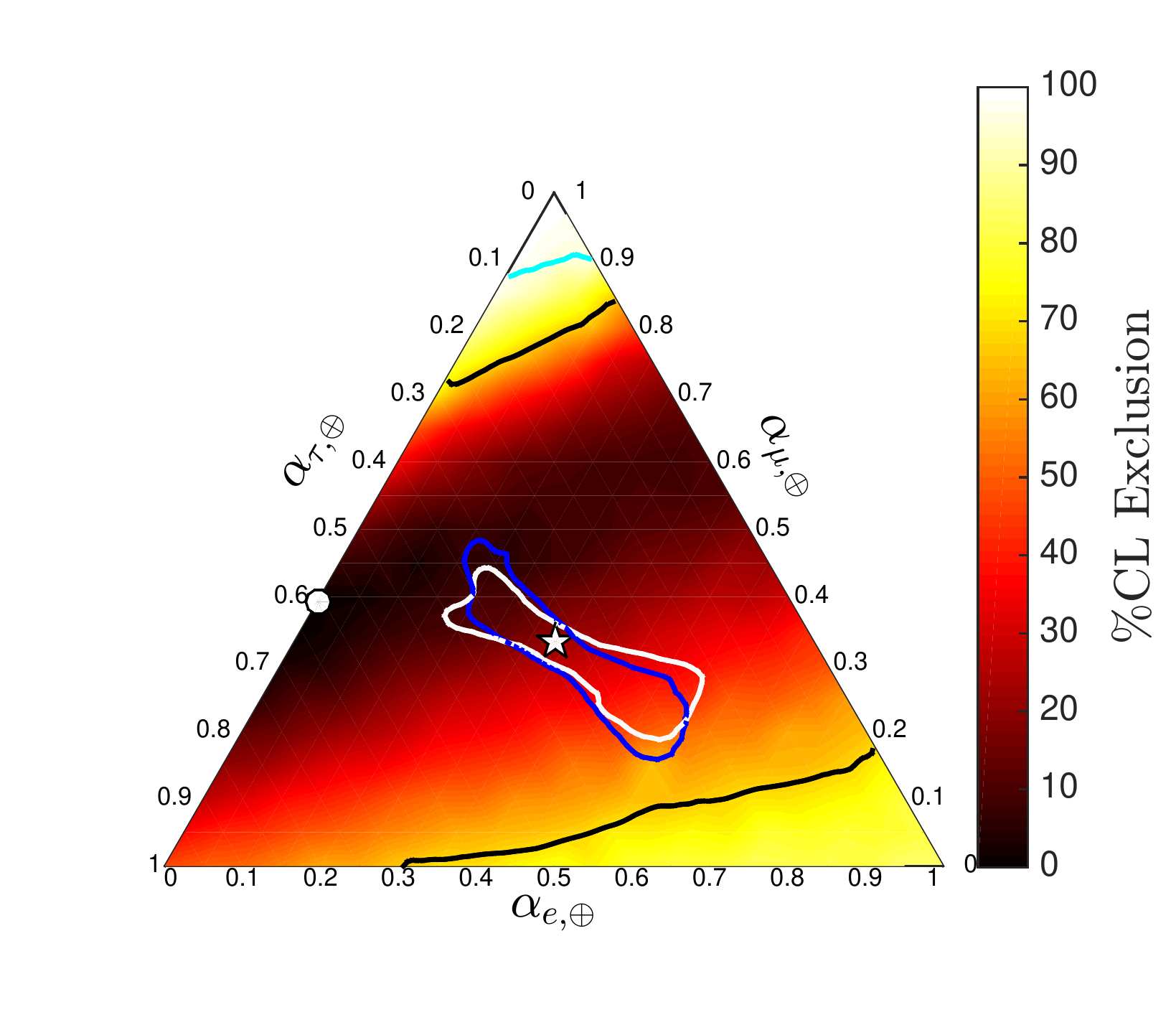} 
	\end{tabular}
	\caption{\textbf{\textit{Isotropic single power-law model: flavor composition.}} Ternary plots of the profile likelihood exclusions of the astrophysical neutrino flavor composition of the high-energy events detected at IceCube after 1347~days, setting the number of prompt atmospheric neutrino events to zero (6P analysis). The black (cyan) lines represent the 68\% (95\%) C.L. allowed regions. We also show the allowed space assuming averaged oscillations during propagation from the sources and taking into account uncertainties at 95\% C.L. of the neutrino mixing angles~\cite{Gonzalez-Garcia:2014bfa}, for normal hierarchy (white contour) and inverted hierarchy (blue contour). The canonical flavor composition at Earth, $(1:1:1)_\oplus$ is also indicated (white star). \textit{Left panel}: using the 53 events in the EM-equivalent deposited energy range [10~TeV $-$ 10~PeV]. Best fit (white circle) at $(0.03 : 0.37 : 0.60)_\oplus$. \textit{Right panel}: using the 32 events in the EM-equivalent deposited energy range [60~TeV $-$ 10~PeV]. Best fit (white circle) at $(0.00 : 0.40 : 0.60)_\oplus$. }
	\label{fig:tern6P}
\end{figure}

In order to compare with previous results obtained with the 3-year data sample and with the preliminary 4-year analysis of the IceCube collaboration, we first consider the simplest case, i.e., an isotropic single power-law flux, which is defined by its flavor composition, normalization and spectral index, and we assume the same properties for neutrinos and antineutrinos. As described above, we perform a 7P fit with $\{\alpha_{e, \oplus}, \alpha_{\mu, \oplus}, \gamma, N_a, N_\mu, N_\nu, N_p\}$ as the set of free parameters, and a 6P fit, which is identical to 7P but fixing $N_p=0$. The results of this section are summarized in Tab.~\ref{tab:1pow}, where we also show the results of the 4P (same as 6P, but fixing the flavor ratio to $(1:1:1)_\oplus$) and 5P (same as 6P, but setting $\alpha_{\mu, \oplus} = \alpha_{\tau, \oplus}$) fits.

\begin{figure}
	\includegraphics[width=0.8\textwidth]{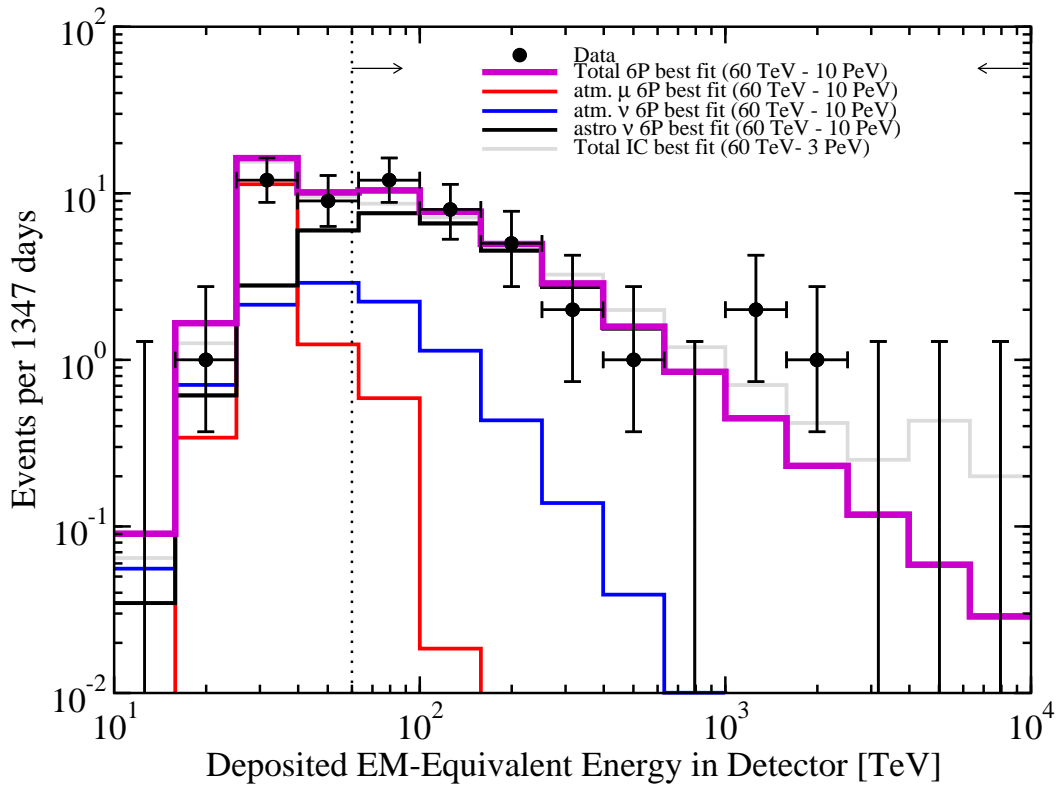} 
	\caption{\textbf{\textit{Isotropic single power-law model: event spectra in the IceCube detector after 1347~days.}} We show the result for the best fit of 6P analysis in the EM-equivalent deposited energy interval [60~TeV $-$ 10~PeV]: atmospheric muon events (red histogram), conventional atmospheric neutrino events (blue histogram), astrophysical neutrino events (black histogram), $E_\nu^2 \, d\Phi/dE_\nu = 10.3 \times 10^{-8} \, (E_\nu/100 \, {\rm TeV})^{-0.77}  \, {\rm GeV}^{-1} \, {\rm cm}^{-2} \, {\rm s}^{-1} \, {\rm sr}^{-1}$ and $(0.00 : 0.40 : 0.60)_\oplus$, and total event spectrum (purple histogram). We also show the spectrum obtained using the preliminary IceCube best fit for $(1:1:1)_\oplus$ in the EM-equivalent deposited energy interval [60~TeV $-$ 3~PeV] (gray histogram), $E_\nu^2 \, d\Phi/dE_\nu = 6.6 \times 10^{-8} \, (E_\nu/100 \, {\rm TeV})^{-0.58}  \, {\rm GeV}^{-1} \, {\rm cm}^{-2} \, {\rm s}^{-1} \, {\rm sr}^{-1}$, and the binned high-energy neutrino event data (black dots)~\cite{Aartsen:2015zva} with Feldman-Cousins errors~\cite{Feldman:1997qc}.}
	\label{fig:events6P}
\end{figure}

In Fig.~\ref{fig:norm6P}, we show our results of the 6P (left panel) and 7P (right panel) fits, considering the data in the two different EM-equivalent deposited energy intervals, [10~TeV $-$ 10~PeV] (filled orange contours) and [60~TeV $-$ 10~PeV] (closed purple curves). The 68\% C.L. (dark colors) and 95\% C.L. (light colors) contours in the $\gamma - \phi$ plane are depicted and we also indicate the best fits (stars). For both 6P and 7P fits, the two C.L. regions are very similar, regardless the energy interval, although they are slightly larger for [60~TeV $-$ 10~PeV], due to the reduced statistics. For the case where the number of events from prompt atmospheric neutrinos is set to zero, 6P fits (left panel), the best fits (bf) agree with each other. For [10~TeV $-$ 10~PeV], we get $\gamma_{\rm bf} = 2.84^{+0.25}_{-0.27}$ and $\phi_{\rm bf} = 11.1^{+3.7}_{-4.8}$ and for [60~TeV $-$ 10~PeV], we get $\gamma_{\rm bf} = 2.77^{+0.31}_{-0.42}$ and $\phi_{\rm bf} = 10.3^{+7.1}_{-6.3}$, where $\phi_{\rm bf}$ is given in the usual units, $10^{-18} \, {\rm GeV}^{-1} \, {\rm cm}^{-2} \, {\rm s}^{-1} \, {\rm sr}^{-1}$. However, when a prompt atmospheric neutrino contribution is also included (7P), the best fit for the spectral index and the flux normalization is slightly larger for the [10~TeV $-$ 10~PeV] data set, yet within the 68\% C.L. regions. Note that these results point to a softer astrophysical spectrum than the 2-year and 3-year best fits~\cite{Aartsen:2013jdh, Aartsen:2014gkd, Palomares-Ruiz:2015mka, Aartsen:2015ivb, Aartsen:2015knd}. They also point to a slightly softer spectrum than the preliminary results presented by the IceCube collaboration using the 4-year data, $\gamma_{\rm IC} = 2.58 \pm 0.25$~\cite{Aartsen:2015zva}, although they are compatible within 1$\sigma$ C.L. However, notice that the preliminary IceCube fit included all events in [60~TeV $-$ 3~PeV], but not the information from the lack of events above 3~PeV. As explained in Ref.~\cite{Palomares-Ruiz:2015mka}, the absence of events near the Glashow resonance~\cite{Glashow:1960zz} ($E_\nu \sim 6.3$~PeV) has implications on the best fit for the flavor composition and the spectral index, either pointing to a suppressed $\bar\nu_e$ flux or to a softening of the spectrum or both. In this sense, the harder spectrum obtained by the IceCube preliminary analysis is expected. Moreover, in our analysis we allow the flavor composition to vary, whereas IceCube's analysis fixes  it to the canonical $(1 : 1 : 1)_\oplus$. Nevertheless, we have checked that this does not have a strong impact on the shape of the spectrum (see the 4P results in Tab.~\ref{tab:1pow}). This can be understood by comparing our 4-year and 3-year results~\cite{Palomares-Ruiz:2015mka}. The larger number of new low-energy events (around $\sim 100$~TeV) relative to the number of high-energy events ($\gtrsim 200$~TeV), steepens the spectrum even further, and thus, the impact of the Glashow resonance is expected to be smaller. Yet, note that both results are compatible at 1$\sigma$ C.L. with the data. On the other hand, recalling that the $\nu_e + \bar\nu_e$ contribution produces a larger number of HESEs than any other flavor in IceCube ({\it cf.} Fig.~12 in Ref.~\cite{Palomares-Ruiz:2015mka}), a larger all-flavor normalization is necessary to explain the same total event rate with a smaller $\nu_e + \bar \nu_e$ relative contribution.

In Fig.~\ref{fig:tern6P} we show the ternary plots of the profile likelihood in flavor space for the two energy intervals we consider. The results are very similar to each other, although the ones for the entire sample (left panel),  [10~TeV $-$ 10~PeV], are slightly more constraining. With the new data, we find a best fit which indicates a slightly larger (smaller) fraction of $\nu_\mu+\bar\nu_\mu$ ($\nu_\tau+\bar\nu_\tau$) in the astrophysical flux than the 3-year results, but likewise with a negligible $\nu_e+\bar\nu_e$ component. With no new event with EM-equivalent deposited energy in the PeV range, we stress again that, assuming an isotropic single-component astrophysical neutrino flux, the lack of events above 2~PeV implies a suppressed $\nu_e + \bar \nu_e$ flux and a steep spectrum (or a break in the spectrum). Concurrently, the important fraction of muon tracks misclassified as showers allows for a significant astrophysical $\nu_\mu+\bar\nu_\mu$ flux. The fact that the best fit for this fraction is higher than in the 3-year analysis can be explained by the higher relative number of tracks: there are 6 new tracks out of 17 new events, whereas there were 8 tracks out of 36 events (plus one coincident event whose energy and direction cannot be reconstructed) in the 3-year data sample. In any case, the canonical $(1 : 1 : 1)_\oplus$ flavor composition is still compatible with the data within $1\sigma$ C.L.

\begin{figure}
	\hspace{-10mm}
	\includegraphics[width=1.05\textwidth]{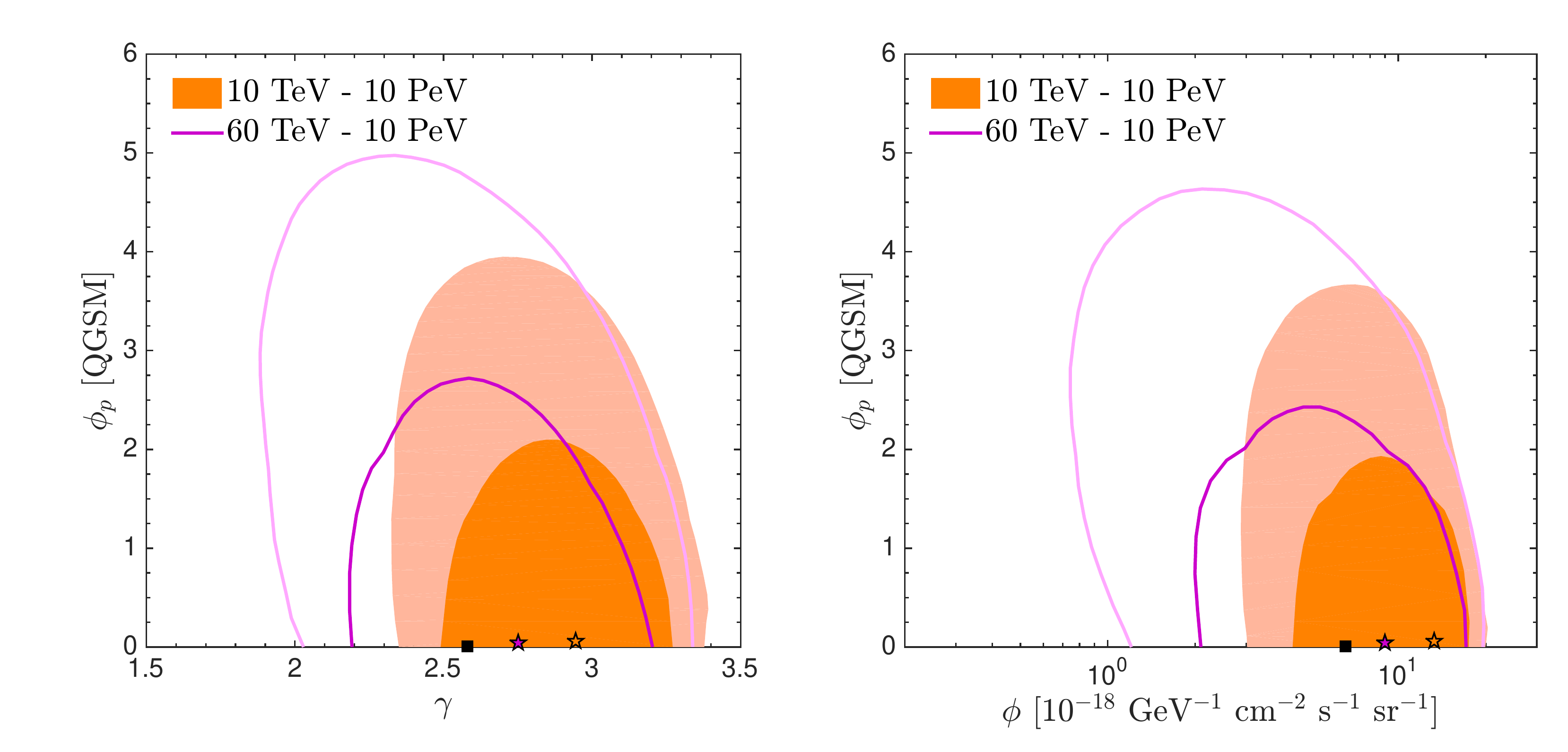} 
	\caption{\textbf{\textit{Isotropic single power-law model:: contribution from prompt atmospheric neutrinos}} versus astrophysical neutrino parameters. Profile likelihood contours at $68\%$ C.L. (dark colors) and $95\%$ C.L. (light colors). The normalization of the prompt contribution, $\phi_{p}$, is quoted with respect to the QGSM prompt flux~\cite{Kaidalov:1984ne, Kaidalov:1986zs, Kaidalov:1985jg, Bugaev:1989we}. Filled orange contours (closed purple curves) represent the EM-equivalent deposited energy interval [10~TeV $-$ 10~PeV] ([60~TeV $-$ 10~PeV]). Our best fits (stars) and the IceCube best fit (square), $\phi_p = 0$, are also indicated. \textit{Left panel:} power-law index of the astrophysical neutrino flux ($\gamma$) - normalization of the prompt atmospheric neutrino ($\phi_{p}$) plane. \textit{Right panel:} normalization of the astrophysical neutrino flux ($\phi$) - normalization of the  prompt atmospheric neutrino ($\phi_{p}$) plane.}
	\label{fig:6Pp}
\end{figure}

In Fig.~\ref{fig:events6P}, we show the 4-year IceCube HESE data (black dots) and the event spectra obtained with our 6P best fit (i.e., setting $N_p = 0$) using events in the EM-equivalent deposited energy interval [60~TeV $-$ 10~PeV]. The astrophysical signal (black histogram) and backgrounds (red and blue histograms), as well as the total (tracks plus showers) spectrum (purple histogram) are depicted. We note that the low energy data is nicely fitted and that our result is in perfect agreement with the preliminary IceCube best fit spectrum (gray histogram) below PeV energies. On the other hand, at PeV energies our best fit spectrum does not present a bump around the Glashow resonance, unlike what happens for the IceCube spectrum. Obviously, this is because $(1:1:1)_\oplus$ is fixed in the IceCube analysis, whereas our best fit in this energy range is $(0.00 : 0.40 : 0.60)_\oplus$, i.e., no $\nu_e+\bar \nu_e$ flux and hence, zero expected Glashow resonance events. However, given the softness of the astrophysical flux, even for the IceCube best fit, the expected number of events around the Glashow resonance after 4 years is less than one.

Finally, we turn to the 7P analysis and discuss the possible contribution from prompt atmospheric neutrinos to the data. In Fig.~\ref{fig:6Pp} we show the 68\% C.L. and 95\% C.L. profile likelihood contours for the normalization of the prompt atmospheric neutrino contribution $\phi_p$, relative to the QGSM flux~\cite{Kaidalov:1984ne, Kaidalov:1986zs, Kaidalov:1985jg, Bugaev:1989we}, versus the spectral index $\gamma$ (left panel) and the normalization $\phi$ (right panel) of the astrophysical neutrino power-law flux. For the two ranges of EM-equivalent deposited energy ([10~TeV $-$ 10~PeV] and [60~TeV $-$ 10~PeV]), the best fit amounts to less than one event in 4 years (0.3 and 0.1), but the result of zero events is well within the 1$\sigma$ C.L. region, which corresponds to a limit of $\sim 1.5$ events per year in the entire energy range, in good agreement with IceCube limits\footnote{Notice, however, that we use QGSM fluxes and the IceCube analyses use the fluxes obtained with the dipole model~\cite{Enberg:2008te, Bhattacharya:2015jpa}.}~\cite{Aartsen:2013eka, Aartsen:2014muf, Aartsen:2015knd}. Recall that we do not impose any prior on the number of prompt atmospheric neutrino events, unlike the preliminary IceCube analysis of the 4-year data~\cite{Aartsen:2015zva}, so we expect our result to be slightly less constraining.

\subsection{Neutrino-antineutrino model}
\label{sec:nunubar}

In standard astrophysical scenarios, ultrahigh-energy neutrinos are produced by the decay of pions and kaons and secondary muons, produced by $pp$ or $p\gamma$ interactions in cosmic accelerators. These different production mechanisms give rise to different neutrino-antineutrino compositions. For instance, in $pp$ sources the same flux of neutrinos and antineutrinos is expected, but in $p\gamma$ sources the flux of neutrinos turns out to be about a factor of two larger, due to the preponderance of $\pi^+$ over $\pi^-$ production. Therefore, it is interesting to check whether the current IceCube data hints at a possible difference in the fluxes of neutrinos and antineutrinos arriving at Earth.

Without further assumptions, given the similarity of the neutrino- and antineutrino-induced event spectra, the only way to be sensitive to a neutrino-antineutrino asymmetry is by using the information provided by the observed event spectrum at deposited energies around the Glashow resonance~\cite{Berezinsky:1977sf, Anchordoqui:2004eb, Bhattacharjee:2005nh, Pakvasa:2007dc, Xing:2011zm, Bhattacharya:2011qu, Barger:2014iua, Palomares-Ruiz:2015mka, Palladino:2015vna, Palladino:2015uoa}. At energies around $E_\nu \simeq 6.3$~PeV, a $\bar \nu_e$ interacting with an electron in the detector volume can produce a $W^-$ boson on resonance with a cross section two orders of magnitude larger than the neutrino-nucleon cross section. Therefore, noting that for $\sim 70\%$ of this type of events (the hadronic modes of $W^-$ decay) most of the neutrino energy is deposited in the detector, a $\bar \nu_e$ flux comparable to the $\nu_e$ flux would result in an excess of events around this deposited energy. So far, the most energetic event in the 4-year HESE sample has an energy of about 2~PeV, far enough from the resonance peak that it is unlikely to be associated with it. The implications of the lack of events around this energy have already been discussed~\cite{Barger:2014iua, Palomares-Ruiz:2015mka, Palladino:2015vna, Palladino:2015uoa}. This could be the indication of a break~\cite{Anchordoqui:2014hua, Palomares-Ruiz:2015mka} or a cutoff~\cite{Learned:2014vya, Winter:2014pya} in the astrophysical neutrino spectrum at an energy of a few PeV.

In this section, we perform a fit to the 4-year HESE data to evaluate the presence of a potential neutrino-antineutrino asymmetry. Assuming the mechanism of production of neutrinos and antineutrinos to be the same, we consider only one power-law spectral index, but allow for different flavor fractions (with $\alpha_{\mu, \oplus} = \alpha_{\tau, \oplus}$ and $\bar \alpha_{\mu, \oplus} = \bar \alpha_{\tau, \oplus}$) and different number of events produced by neutrinos and antineutrinos. As explained above, we perform a 7P+$\nu\bar \nu$ fit, where the set of free parameters is $\{\alpha_{e, \oplus}, \bar\alpha_{e, \oplus}, \gamma, N_a, \bar N_a, N_\mu, N_\nu \}$. The results of this section are summarized in Tab.~\ref{tab:nunubar}.

\begin{figure}
	\includegraphics[width=0.7\textwidth]{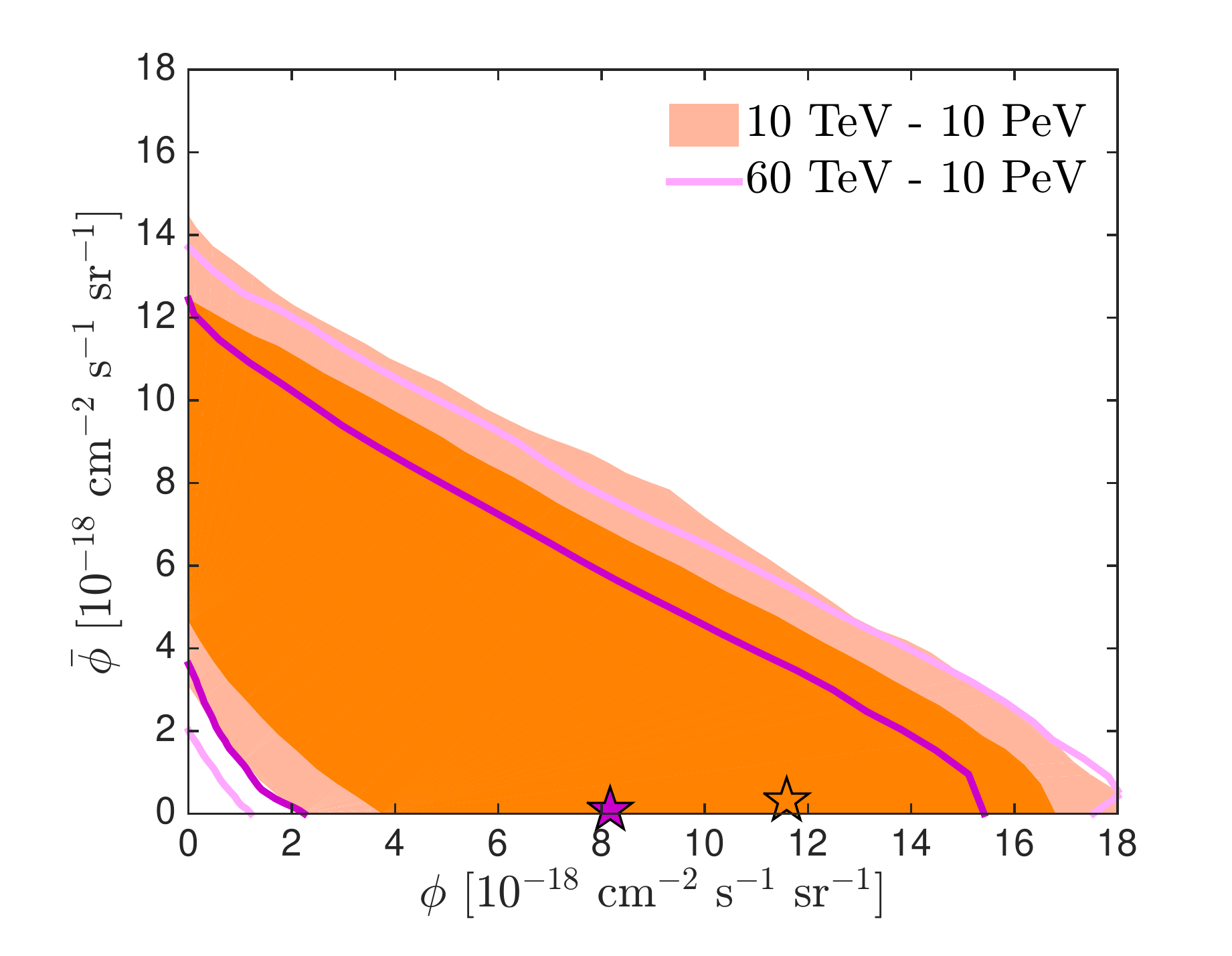} 
	\caption{\textbf{\textit{Neutrino-antineutrino model: normalization of the fluxes.}} Profile likelihood contours in the $\phi - \bar \phi$ plane, at 68\% C.L. (dark colors) and 95\% C.L. (light colors) with the 7P+$\nu\bar \nu$ analysis, as described in the text. Filled orange contours (closed purple curves) represent the EM-equivalent deposited energy interval [10~TeV $-$ 10~PeV] ([60~TeV $-$ 10~PeV]). The best fits (stars) are also indicated.}
	\label{fig:normnunubar}
\end{figure}

In contrast with the 6P analysis, in which the sum of neutrino and antineutrino events is fitted, we now find the number of astrophysical neutrino or antineutrino events not to be very correlated with the spectral index of the astrophysical flux (assumed to be the same for both populations). Unsurprisingly, the numbers of neutrino and antineutrino events are highly correlated between themselves, to keep the total number of events consistent with the data. This is an indication of the degeneracy in both, the normalization of the fluxes and the flavor ratios of neutrinos and antineutrinos\footnote{See Ref.~\cite{Nunokawa:2016pop} for a discussion of the phenomenological implications of an asymmetry in the neutrino and antineutrino flavor ratios.}. This is shown in Fig.~\ref{fig:normnunubar}, where we depict the 68\% C.L. and 95\% C.L. profile likelihood contours in the $\phi - \bar \phi$ plane for the two energy intervals. Solutions with either no astrophysical neutrino or no astrophysical antineutrino flux are allowed within 1$\sigma$ C.L., even though the best fit corresponds to a neutrino-dominated flux. This implies that current data do not allow us to determine if there is an asymmetry in the neutrino-antineutrino composition of the astrophysical flux and of course, do not allow us to distinguish scenarios with different fractions of neutrinos and antineutrinos (see also Refs.~\cite{Barger:2014iua, Palladino:2015uoa}).

\section{Two power-law analysis}
\label{sec:2powNS}

The case of a single isotropic high-energy neutrino flux is the simplest scenario one could consider. However, different astrophysical sources are likely to contribute to the neutrino flux arriving at Earth and thus, it is also natural to consider more complicated spectral features and anisotropies in the angular distribution. As the next step, we discuss the possibility of a two-component astrophysical flux. This has already been suggested to explain the IceCube HESE spectrum, either assuming an isotropic flux and with a focus on the gap below 1~PeV~\cite{Chen:2014gxa} or considering galactic (mainly in the southern hemisphere) and extragalactic (isotropic) contributions with different (but fixed) spectra~\cite{ Palladino:2016zoe}, but with the same flavor composition. Here, we update and extend these analyses. Firstly, we consider an isotropic model with two components with different energy spectra and flavor compositions. Secondly, we turn to the case of a different (single) power-law flux from the northern and southern hemispheres. Unlike the plots in the previous section, we show credible regions using a bayesian analysis of our posterior distributions. We do so because for this larger parameter space, robust profile likelihoods require impractically large samples. The results of this section are summarized in Tabs.~\ref{tab:2pow} and~\ref{tab:NS}, where we also quote the normalization of the fluxes ($\phi$), instead of the total number of events ($N$).

\begin{figure}
	\begin{tabular}{l l}
		\hspace{-5mm}
		\includegraphics[width=0.55\textwidth]{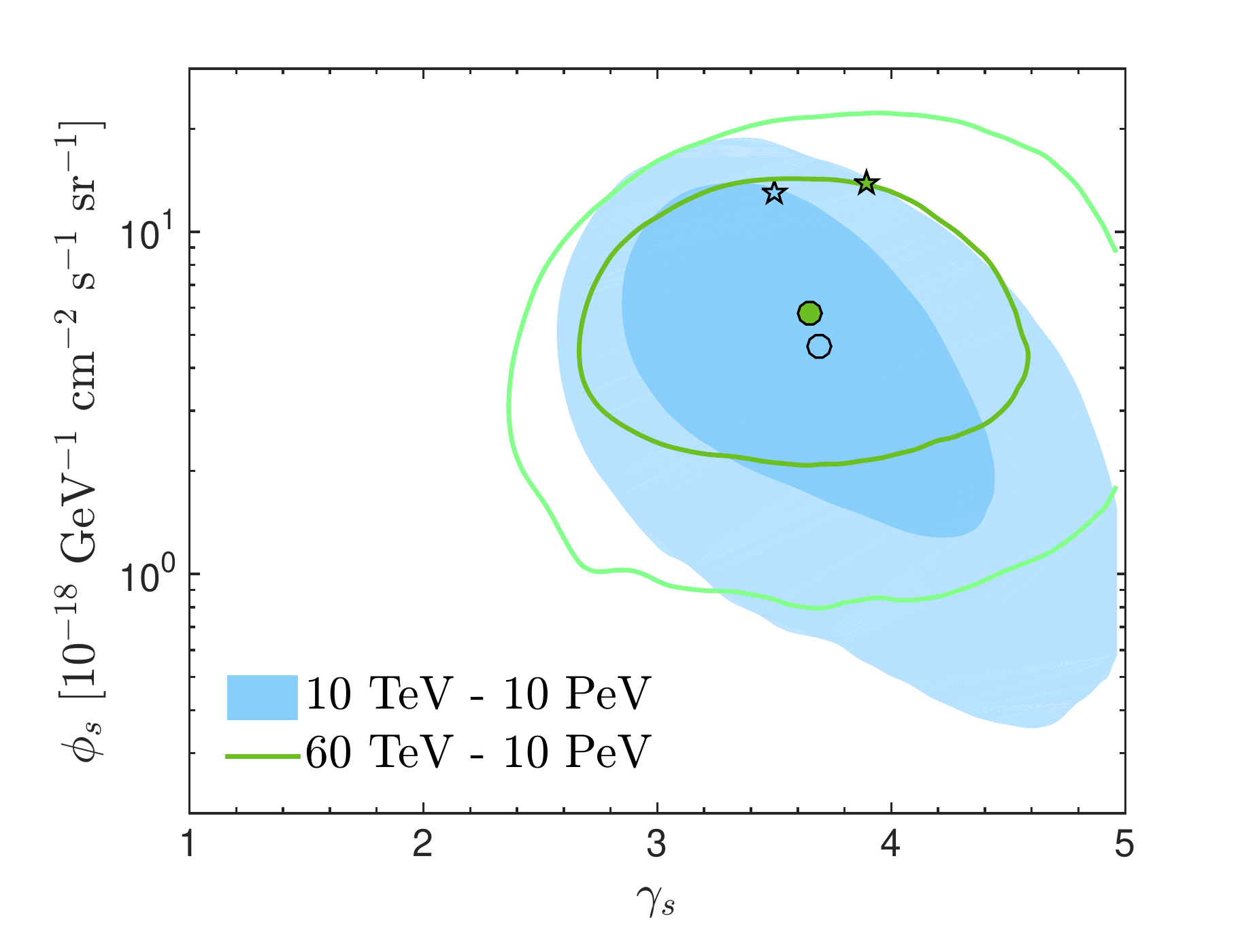} & \hspace{-10mm} \includegraphics[width=0.55\textwidth]{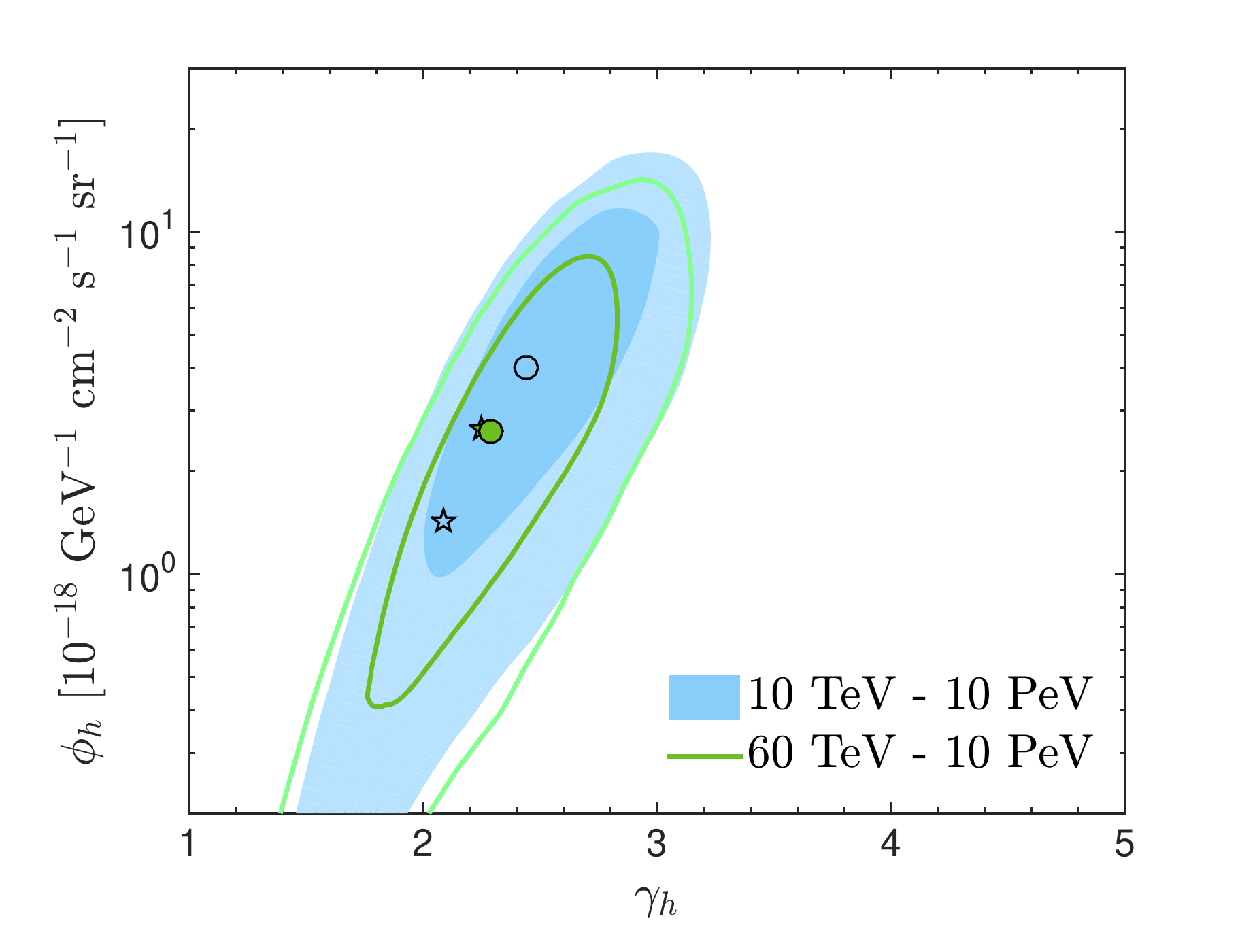} 
	\end{tabular}
	\caption{\textbf{\textit{Isotropic two power-law model: spectral shape.}} Contours in the $\gamma - \phi$ plane, corresponding to 68\% (dark colors) and 95\% (light colors) credible regions with the 8P+2pow analysis, as described in the text. Filled blue contours (closed green curves) represent the EM-equivalent deposited energy interval of [10~TeV $-$ 10~PeV] ([60~TeV $-$ 10~PeV]). The posterior means (circles) and best fits (stars) are also indicated. \textit{Left panel}: astrophysical component with a soft spectrum. \textit{Right panel:} astrophysical component with a hard spectrum.}
	\label{fig:norm2pow}
\end{figure}

\subsection{Isotropic two-power-law model}
\label{sec:2pow}

We first consider the simplest extension of an isotropic power-law flux, i.e., an isotropic flux with two power-law components. We vary the flavor composition (with $(\alpha_{\mu, \rm s})_\oplus = (\alpha_{\tau, \rm s})_\oplus$ and $(\alpha_{\mu, \rm h})_\oplus = (\alpha_{\tau, \rm h})_\oplus$), the power-law indices and the number of events produced by each neutrino flux component. As in the single component case, we assume the neutrino and antineutrino fluxes to be equal. As described above, we perform an 8P+2pow fit with the free parameters: $\{(\alpha_{e, \rm s})_\oplus, (\alpha_{e, \rm h})_\oplus, \gamma_{\rm s}, \gamma_{\rm h}, N_{a, \rm s}, N_{a, \rm h}, N_\nu, N_\mu \}$, where the indices `s' and `h' refer to the soft and hard component, respectively. The results of this section are summarized in Tab.~\ref{tab:2pow}.

In Fig.~\ref{fig:norm2pow} we show the results of the 68\% and 95\% credible regions in the $\gamma_{\rm s} - \phi_{\rm s}$ (left panel) and $\gamma_{\rm h} - \phi_{\rm h}$ (right panel) planes. In both panels, in addition to the best fits (stars) we also show the posterior means (circles). We see that for the component that would explain the low-energy data, the best fit spectrum is very soft, $(\gamma_{\rm s})_{\rm bf} = 3.50^{+1.55}_{-0.41}$ for [10~TeV $-$ 10~PeV] and $(\gamma_{\rm s})_{\rm bf} = 3.89^{+1.08}_{-0.16}$ for [60~TeV $-$ 10~PeV], with a spectrum similar in shape to the conventional atmospheric neutrino flux. This partial degeneracy explains the low number of atmospheric neutrino events obtained from the fit, mainly when the entire energy range is considered.  We also note that the best fits for the normalization of the soft component are at the edge of the 68\% credible regions, with the posterior means being about a factor of 2 smaller. On the other hand, the best fits for the power-law index of the hard component, $(\gamma_{\rm h})_{\rm bf} = 2.09^{+0.92}_{-0.64}$ for the interval [10~TeV $-$ 10~PeV] and $(\gamma_{\rm h})_{\rm bf} = 2.25^{+0.58}_{-0.81}$ for [60~TeV $-$ 10~PeV], are in much better agreement with the through-going muon sample~\cite{Aartsen:2015knd, Aartsen:2015rwa}. Note that also the all-flavor normalization obtained for the hard component is in agreement with the normalization of the $\nu_\mu + \bar \nu_\mu$ flux that best fits the through-going muon data. The best fit and the posterior mean of the hard component are almost identical for the [60~TeV $-$ 10~PeV] interval.

\begin{figure}
	\includegraphics[width=0.7\textwidth]{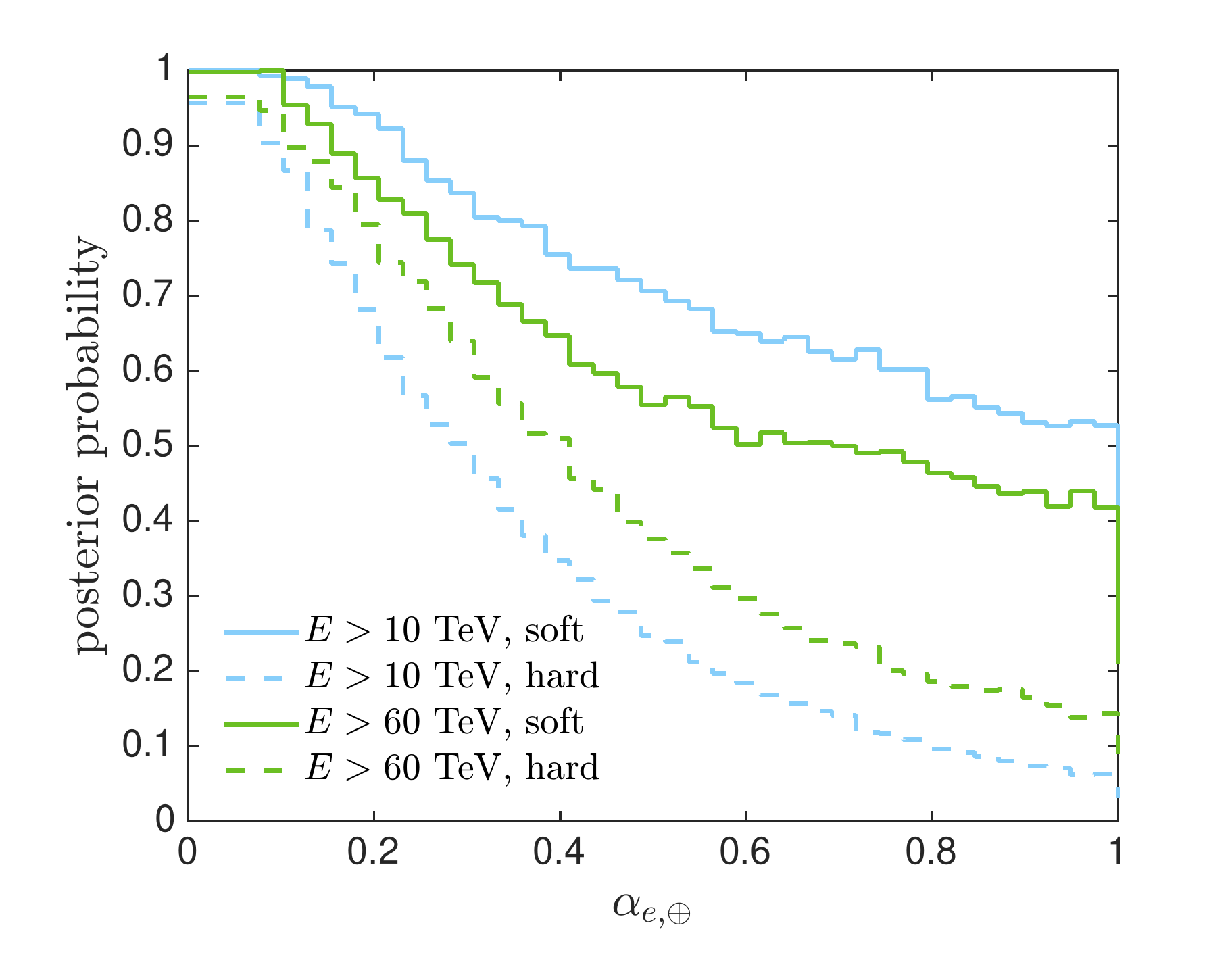} 
	\caption{\textbf{\textit{Isotropic two power-law model: flavor composition.}} Posterior probability distributions of the $\nu_e + \bar \nu_e$ fractions for the soft ($(\alpha_{e,\rm s})_\oplus$, solid upper curves) and hard   ($(\alpha_{e,\rm h})_\oplus$, dashed lower curves) components of the astrophysical neutrino flux, assuming $(\alpha_{\mu,\rm s})_\oplus = (\alpha_{\tau,\rm s})_\oplus$ and $(\alpha_{\mu,\rm h})_\oplus = (\alpha_{\tau,\rm h})_\oplus$. We show the results of the analyses of the two intervals in EM-equivalent deposited energy considered in this work: [10~TeV $-$ 10~PeV] (blue curves) and [60~TeV $-$ 10~PeV] (green curves).}
	\label{fig:alphas2pow}
\end{figure}

In Fig.~\ref{fig:alphas2pow}, the posterior probability distributions of the flavor compositions of both flux components are presented. As we are assuming equal $\nu_\mu + \bar \nu_\mu$ and $\nu_\tau + \bar \nu_\tau$ fractions, we only show the results for the $\nu_e + \bar \nu_e$ component, for the usual energy intervals. Similarly to the single power-law analysis (6P), the case of a negligible $\nu_e + \bar \nu_e$ fraction turns out to be the highest probability point for both components, although the hard component is more peaked towards $(\alpha_{e,\rm h})_\oplus = 0$ than the soft component, for which the distribution is flatter. Again, this is due to the lack of events around the Glashow resonance, which is more important for a hard spectrum. Thus, for an unbroken power-law spectrum, the only way of reducing the importance of the excess of events around these energies ($\sim 6.3$~PeV), without significantly changing the expected number of events at lower energies, is by means of a $\bar\nu_e$ flux which is suppressed with respect to the $\nu_\mu+\bar\nu_\mu$ and $\nu_\tau+\bar\nu_\tau$ fluxes.

\begin{figure}
	\includegraphics[width=0.8\textwidth]{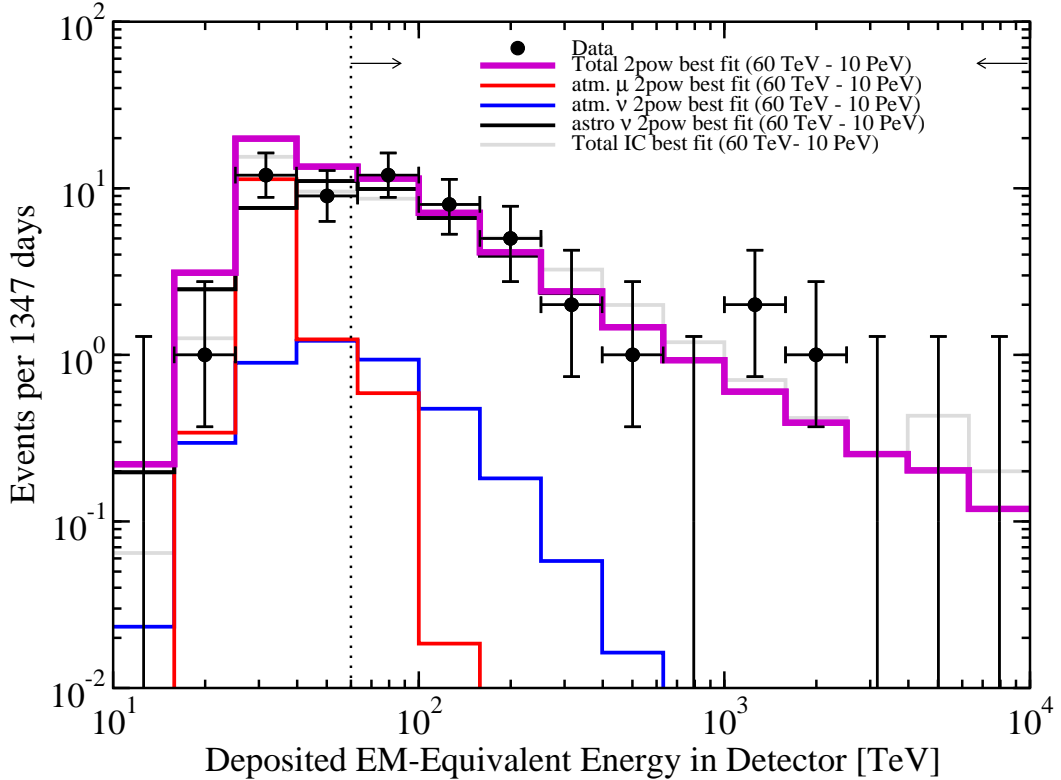} 
	\caption{\textbf{\textit{Isotropic two power-law model: event spectra in the IceCube detector after 1347~days.}} We show the result of the 8P+2pow best fit in the EM-equivalent deposited energy interval [60~TeV $-$ 10~PeV]: atmospheric muon events (red histogram), conventional atmospheric neutrino events (blue histogram), astrophysical neutrino events (black histogram), $E_\nu^2 \, d\Phi/dE_\nu = [ 13.8 \, (E_\nu/100 \, {\rm TeV})^{-1.89} + 2.7 \, (E_\nu/100 \, {\rm TeV})^{-0.25} ]  \times 10^{-8} \, {\rm GeV}^{-1} \, {\rm cm}^{-2} \, {\rm s}^{-1} \, {\rm sr}^{-1}$ with $(\alpha_{e ,\rm s})_\oplus = 0.01$ and $(\alpha_{e ,\rm h})_\oplus = 0.03$, and total event spectrum (purple histogram). We also show the spectrum obtained using the preliminary IceCube best fit  for $(1:1:1)_\oplus$ in the EM-equivalent deposited energy interval [60~TeV $-$ 3~PeV] (gray histogram), $E_\nu^2 \, d\Phi/dE_\nu = 6.6 \times 10^{-8} \, (E_\nu/100 \, {\rm TeV})^{-0.58}  \, {\rm GeV}^{-1} \, {\rm cm}^{-2} \, {\rm s}^{-1} \, {\rm sr}^{-1}$, and the binned high-energy neutrino event data (black dots)~\cite{Aartsen:2015zva} with Feldman-Cousins errors~\cite{Feldman:1997qc}.}
	\label{fig:events2pow}
\end{figure}

Finally, in Fig.~\ref{fig:events2pow} we show the 4-year IceCube HESE data (black dots) and the event spectra corresponding to the best fit in the EM-equivalent deposited energy interval [60~TeV $-$ 10~PeV] of the two power-law model for the astrophysical neutrino flux. As in Fig.~\ref{fig:events6P}, the signal (black histogram) and backgrounds (red and blue histograms), as well as the total spectrum (purple histogram) are depicted. Our best fit for the total (tracks plus showers) event spectrum turns out to be very close to the IceCube result for the single power-law model, using the EM-equivalent deposited energy interval [60~TeV $-$ 3~PeV] and assuming $(1:1:1)_\oplus$ (gray histogram). Again, with a suppressed $\bar \nu_e$ contribution, no bump appears in our best fit around the Glashow resonance. 

At this point, it is interesting to quantify if there is a statistically significant preference for the two power-law model over the single power-law, assuming isotropy. In order to do this, we compare the best fits obtained for the single power-law model with the 4P, with $(\alpha_e : \alpha_\mu : \alpha_\tau)_\oplus = (1:1:1)_\oplus$, and 5P analyses, with $(\alpha_e : \alpha_\mu : \alpha_\tau)_\oplus = (\alpha_e : (1-\alpha_e)/2 : (1-\alpha_e)/2)_\oplus$, and that obtained with the 8P+2pow analysis (see Tab.~\ref{tab:2pow}). The values for the log-likelihood ratios and the $p$-values\footnote{Note that these log-likelihood ratios do not follow a $\chi^2$ distribution, but have larger expectation values. Therefore, the quoted numbers represent lower bounds on the actual $p$-values.} corresponding to 4 and 3 degrees of freedom (dof), for the two energy intervals, are:
\begin{eqnarray}
-2 \ln\left({\frac{{\cal L}_{\rm max}(\rm 4P)}{{\cal L}_{\rm max}(\rm 8P+2pow)}}\right)_{[10~{\rm TeV}-10~{\rm PeV}]} & = & 2.82 \hspace{5mm} (p\textrm{-value=0.59 for 4 dof}) \\
-2 \ln\left({\frac{{\cal L}_{\rm max}(\rm 5P)}{{\cal L}_{\rm max}(\rm 8P+2pow)}}\right)_{[10~{\rm TeV}-10~{\rm PeV}]} & = & 1.72 \hspace{5mm} (p\textrm{-value=0.63 for 3 dof}) \\
-2 \ln\left({\frac{{\cal L}_{\rm max}(\rm 4P)}{{\cal L}_{\rm max}(\rm 8P+2pow)}}\right)_{[60~{\rm TeV}-10~{\rm PeV}]} & = & 1.46 \hspace{5mm} (p\textrm{-value=0.83 for 4 dof}) \\
-2 \ln\left({\frac{{\cal L}_{\rm max}(\rm 5P)}{{\cal L}_{\rm max}(\rm 8P+2pow)}}\right)_{[60~{\rm TeV}-10~{\rm PeV}]} & = & 1.05 \hspace{5mm} (p\textrm{-value=0.79 for 3 dof})
\end{eqnarray}
Therefore, with only HESE data, currently it is not possible to distinguish between a single power-law astrophysical neutrino flux and a two-component flux, assuming isotropy, as the differences are at less than 1$\sigma$ C.L.

\subsection{North-South model}
\label{sec:NS}

A recent analysis by the IceCube collaboration using different data samples~\cite{Aartsen:2015knd}, including through-going track events from the northern hemisphere, concluded that there is a hint of a North-South asymmetry in the shape of the astrophysical spectrum with a 1.1$\sigma$ discrepancy with respect to the isotropic single power-law model. This would be particularly interesting as it could be an indication of a galactic contribution in addition to an isotropic extragalactic flux\footnote{Even if a flux with galatic plus extragalactic contributions would approximately correspond to a single-component flux from the North and a two-component flux from the South, we have shown in the previous section (Sec.~\ref{sec:2pow}) that the two power-law analysis does not significantly improve the fit to the current HESE data over the single power-law model. Thus, this scenario can be adequately modeled by the North-South model considered here.}, which has already been suggested using different arguments~\cite{Neronov:2013lza, Razzaque:2013uoa, Neronov:2015osa, Palladino:2016zoe, Neronov:2016bnp}. 

Here, we would like to examine whether the HESE data are partly driving this asymmetry, using an extra year of data with respect to Ref.~\cite{Aartsen:2015knd}. We therefore examine the statistical significance of the North-South asymmetry by using the 4-year HESE data and performing an 8P+NS analysis, where the set of free parameters is $\{(\alpha_{e, \rm N})_\oplus, (\alpha_{e, \rm S})_\oplus, \gamma_{\rm N}, \gamma_{\rm S}, N_{a, \rm N}, N_{a, \rm S}, N_\mu, N_\nu \}$, as described above. The results of this section are summarized in Tab.~\ref{tab:NS}.

\begin{figure}
	\begin{tabular}{l l}
		\hspace{-5mm}
		\includegraphics[width=0.55\textwidth]{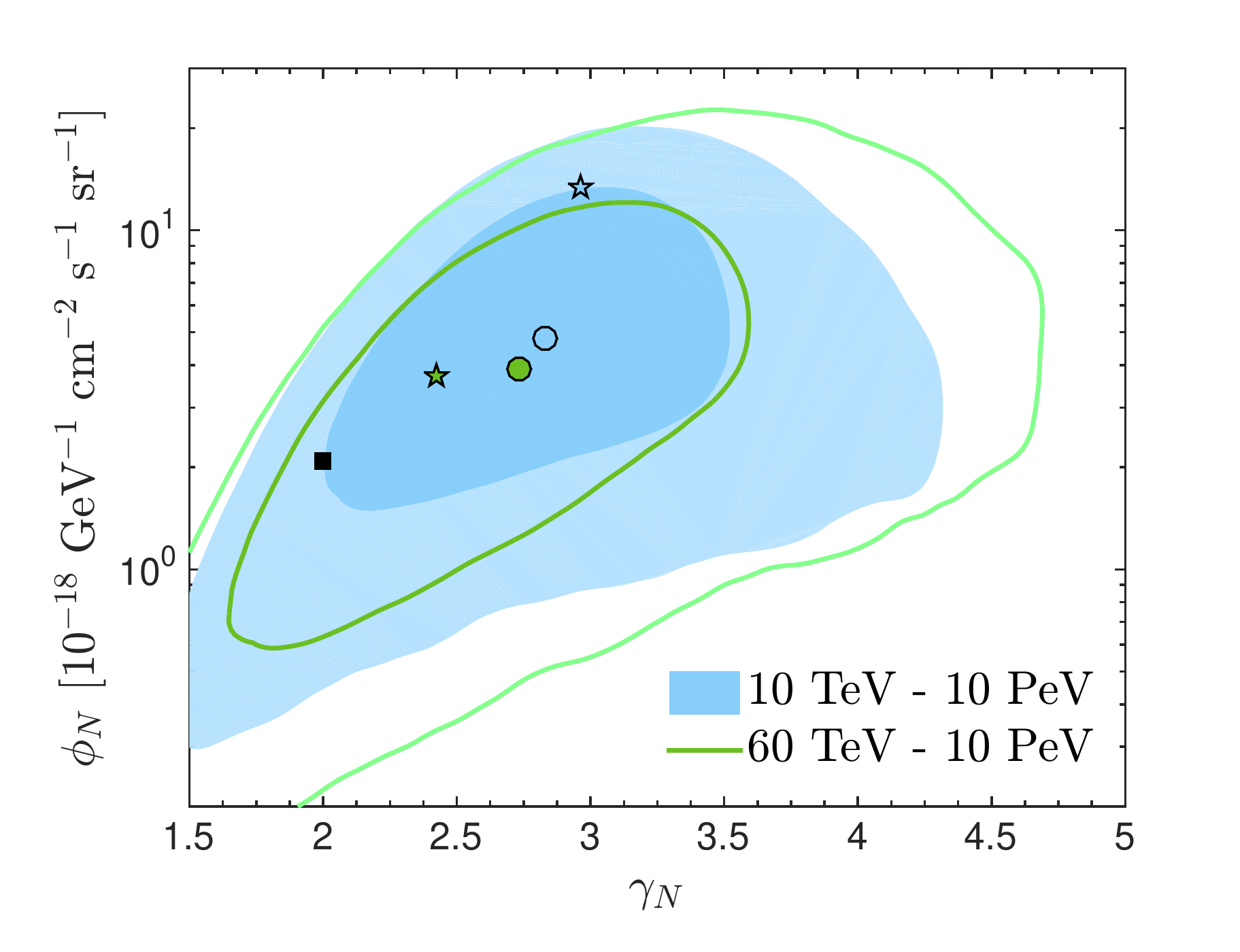} & \hspace{-10mm} \includegraphics[width=0.55\textwidth]{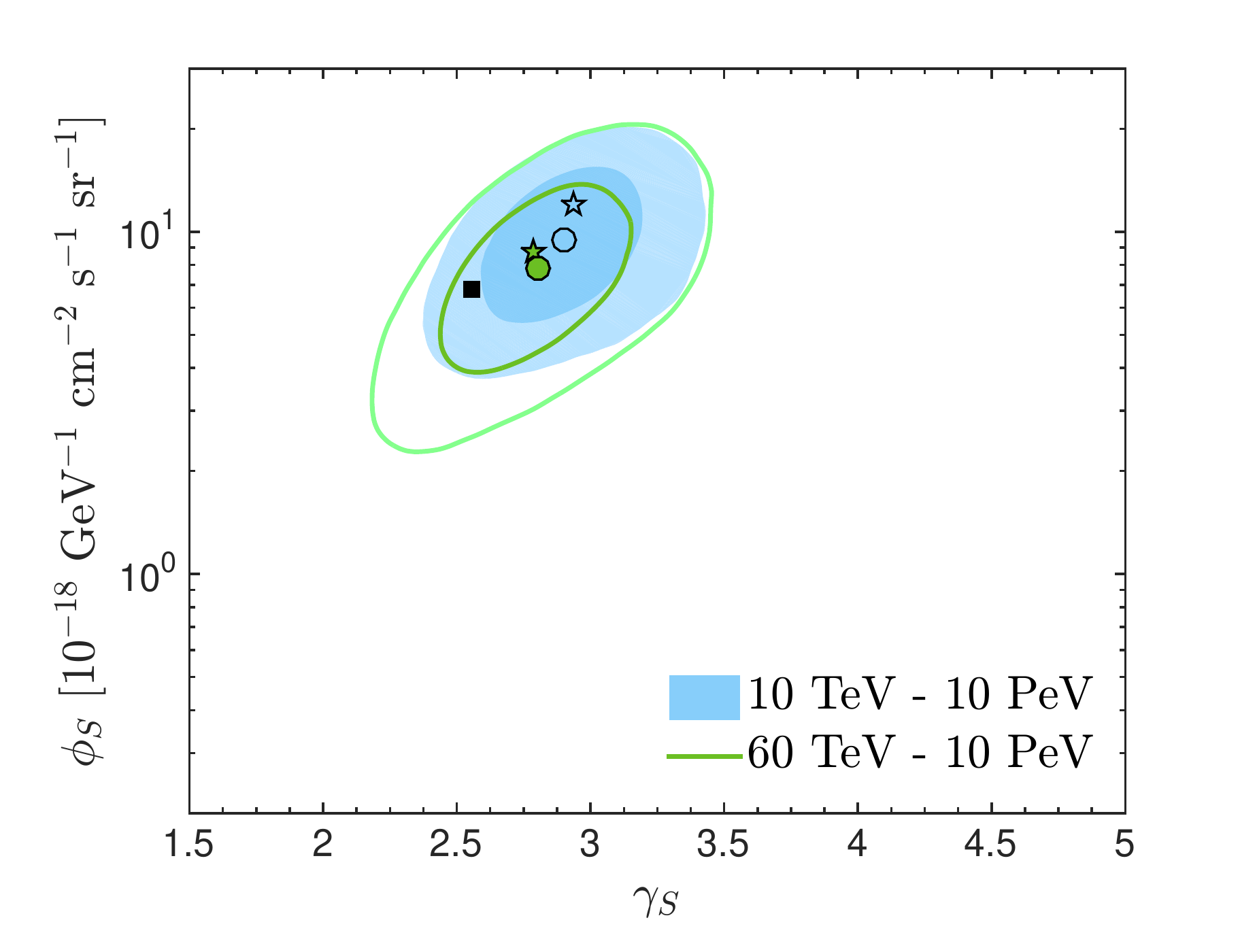} 
	\end{tabular}
	\caption{\textbf{\textit{North-South model: spectral shape.}} Contours in the $\gamma - \phi$ plane, corresponding to 68\% (dark colors) and 95\% (light colors) credible regions with the 8P+2pow analysis, as described in the text. Filled blue contours (closed green curves) represent the EM-equivalent deposited energy interval of [10~TeV $-$ 10~PeV] ([60~TeV $-$ 10~PeV]). The posterior means (circles) and best fits (stars) are also indicated, as well as the IceCube best fit of a similar analysis with a larger data  sample~\cite{Aartsen:2015knd} (squares). \textit{Left panel}: upgoing events (northern hemisphere). \textit{Right panel:} downgoing events (southern hemisphere).}
	\label{fig:normupdn}
\end{figure}

The results for the power-law indices and normalizations of the astrophysical spectrum are shown in Fig.~\ref{fig:normupdn} for neutrinos originating from the northern (left panel) and southern (right panel)  hemispheres. There is clearly no indication of an asymmetry in the shape of the spectrum. In the case of the EM-equivalent deposited energy interval [10~TeV $-$ 10~PeV], the best fit values for the spectral indices are $(\gamma_{\rm N})_{\rm bf} = 2.96^{+0.41}_{-0.78}$ and $(\gamma_{\rm S})_{\rm bf} = 2.94^{+0.24}_{-0.29}$, which are in perfect agreement with each other. The best fits for the spectral indices in the interval [60~TeV $-$ 10~PeV] are: $(\gamma_{\rm N})_{\rm bf} = 2.42^{+0.89}_{-0.61}$ and $(\gamma_{\rm S})_{\rm bf} = 2.79^{+0.31}_{-0.29}$, and both spectra are compatible at $1\sigma$ C.L., mainly due to the large uncertainty  from the upgoing, northern hemisphere, events. Regardless the energy interval considered, the best fit for the southern hemisphere is very similar to what is obtained for the single power-law isotropic model (see Tab.~\ref{tab:1pow}). This is not surprising, as the statistics are dominated by the downgoing events. Indeed, the scarce data from the northern hemisphere also explains the large allowed regions and the very small dependence on the deposited energy interval which is considered. Note that in the interval [10~TeV $-$ 10~PeV] ([60~TeV $-$ 10~PeV]), only 16 (10) events have been observed from the northern hemisphere. The asymmetry found in Ref.~\cite{Aartsen:2015knd} is thus likely dominated by the addition of through-going muon track events, which entirely originate from the northern hemisphere. These events are well-fitted by a much harder spectrum, $\gamma_{\rm TG} \simeq 2$~\cite{Aartsen:2015rwa, Aartsen:2015zva}. When combined with the mainly-downgoing, much softer HESE spectrum, this results in the reported asymmetry.

It is important to stress that we are not excluding the possibility of the quoted asymmetry, but trying to understand its origin. Indeed, our results for the northern sky are compatible at the 1$\sigma$ C.L. with an astrophysical spectrum as hard as the one found from the analysis of the through-going track sample. The sample size is simply not large enough to make stronger claims with the HESE data set on its own. Furthermore, the asymmetry could occur in the normalization, rather than in the spectral index. Nevertheless, our results do not show either the presence of any asymmetry in this parameter.

It is also interesting to comment on the flavor composition resulting from this analysis. In the EM-equivalent deposited energy interval [10~TeV $-$ 10~PeV] (analogous arguments can be drawn for the  [60~TeV $-$ 10~PeV] case), there are 9~track events from the southern hemisphere. However,  most of them are likely produced by atmospheric muons\footnote{However, note that event 45 is a downgoing track with a deposited energy of 430~TeV, which is very likely of astrophysical origin.}, which would imply a suppression of the $\nu_\mu+\bar\nu_\mu$ flux.  Because we are fixing the $\nu_\mu+\bar\nu_\mu$ flux to be equal to the $\nu_\tau+\bar\nu_\tau$ flux, this also implies a suppressed $\nu_\tau+\bar\nu_\tau$ flux. However, the lack of events around the Glashow resonance also implies a suppressed $\nu_e+\bar\nu_e$ flux. These two features, along with the condition $(\alpha_{\mu,\rm S})_\oplus = (\alpha_{\tau,\rm S})_\oplus$, are in tension and this finally results in a significant contribution of  $\nu_e+\bar\nu_e$ to the total astrophysical neutrino flux, $(\alpha_{e, \rm S})_{\oplus, \rm bf}=0.26$ and $(\alpha_{e, \rm S})_{\oplus, \rm bf}=0.28$ for [10~TeV $-$ 10~PeV] and [60~TeV $-$ 10~PeV], respectively. On the other hand, there are 5~track events from the northern hemisphere that are most likely of astrophysical origin. This implies that a significant $\nu_\mu+\bar\nu_\mu$ flux is expected and thus, an equally high $\nu_\tau+\bar\nu_\tau$ flux, resulting in a suppressed $\nu_e+\bar\nu_e$ flux. This indicates a small North-South asymmetry in the flavor composition, although given the statistics and the large uncertainties in the backgrounds, it is not significant at all.

\section{Summary}
\label{sec:summary}

After four years of data taking, the IceCube neutrino telescope has detected 53 high-energy neutrino events with EM-equivalent deposited energies between 20~TeV and 2~PeV (plus one event whose energy and direction cannot be unambiguously reconstructed). This is a breakthrough towards our understanding of the most violent processes of the Universe and opens the door to study the most powerful cosmic sources in a completely new manner, complementary to $\gamma$-ray and cosmic-ray observations. 

In previous works, we studied the flavor composition of the astrophysical neutrino flux arriving at Earth~\cite{Mena:2014sja, Palomares-Ruiz:2014zra} by performing a simplified analysis of the 2- and 3-year HESE data. We concluded that the best fit was obtained for a flavor ratio of  $(1:0:0)_\oplus$, but remained compatible with the canonical $(1:1:1)_\oplus$ at 2$\sigma$ C.L. Nevertheless, this analysis did not include spectral or directional information, and made some simplifying approximations. More importantly, it did not include the information from the lack of events around the Glashow resonance nor the fact that, for HESE data, 30\% of tracks were expected to be misclassified as showers, which was not disclosed until the publication of Ref.~\cite{Aartsen:2015ivb}. In Ref.~\cite{Palomares-Ruiz:2015mka} (see also Ref.~\cite{Vincent:2015woa}), we performed a spectral analysis of the 3-year HESE data where the flavor composition of the astrophysical neutrino flux, its spectral index and normalization, along with the number of background events were left free to vary and where we also considered the misclassification of tracks as showers and the implications of the non-detection of events above 2~PeV. This analysis improved upon previous ones in the literature where either the flavor composition was held fixed~\cite{Aartsen:2013jdh, Aartsen:2014gkd, Chen:2013dza, Kalashev:2014vra} or the backgrounds were not allowed to vary~\cite{Watanabe:2014qua}. It reached different conclusions to the ones in Refs.~\cite{Mena:2014sja, Palomares-Ruiz:2014zra} regarding the flavor composition, with a best fit for $\nu_\tau+\bar\nu_\tau$ dominance and with $(1:1:1)_\oplus$ compatible with data at 1$\sigma$ C.L. These results are mainly explained by considering a larger deposited energy interval which includes the Glashow resonance and a proper treatment of track misidentification. These findings were later confirmed~\cite{Aartsen:2015ivb, Aartsen:2015knd}.

In this work, we have studied the IceCube HESE sample with an extra year of data, which includes 17 new events. We have refined and updated our previous analyses in a number of ways. Namely, we have improved the calculation of the veto for downgoing atmospheric neutrinos and the modeling of the atmospheric muon background (Sec.~\ref{sec:bkg}), the calculation of the energy resolution has also been revised (Sec.~\ref{sec:res}) and a number of different possible scenarios has been scrutinized (Sec.~\ref{sec:likelihood} for a description of the likelihoods), including the usual isotropic unbroken power-law spectrum, the possibility of a neutrino-antineutrino asymmetry, two-component fluxes and a North-South model. All the fits have been performed using the detected events in two EM-equivalent deposited energy intervals: one with the usual lower limit quoted by the IceCube collaboration, but with a higher upper limit than in some IceCube analyses, [60~TeV $-$ 10~PeV], and a second one using the entire data sample, [10~TeV $-$ 10~PeV]. Our results are given numerically in Tabs.~\ref{tab:1pow}$-$\ref{tab:NS} and can be summarized as follows:

\begin{itemize}
 \item In Sec.~\ref{sec:1pow}, we have first considered the simplest scenario for the astrophysical neutrino flux, i.e., an isotropic unbroken power-law spectrum, and we have updated the results with 4 years of HESE data. The additional events at low energies and the non-detection of events in the PeV range result in a softer spectrum than what the 3-year data suggested, in agreement with the preliminary IceCube analysis~\cite{Aartsen:2015zva}. We also find a weaker dependence on the energy interval, plausibly due to increased statistics filling in for previous fluctuations. The canonical $(1:1:1)_\oplus$ flavor ratio provides a very good fit to the event spectrum, although the best fit still implies a paucity of $\nu_e+\bar\nu_e$ . The flavor composition is correlated with the shape of the spectrum and any model that lowers the expected flux around the Glashow resonance, such as a break in the power law at a few PeV, could move the best fit back towards $(1:1:1)_\oplus$~\cite{Palomares-Ruiz:2015mka, Aartsen:2015knd}. Finally, we have also included a possible contribution from a prompt atmospheric neutrino component, but we obtain a result compatible with zero events and a limit in good agreement with IceCube results~\cite{Aartsen:2013eka, Aartsen:2014muf, Aartsen:2015knd}. 

\item In Sec.~\ref{sec:nunubar} we have evaluated the possibility of a neutrino-antineutrino asymmetry, which could naturally occur in sources where high-energy neutrinos are produced via the interaction of cosmic-rays with the radiation background field (via $p\gamma$ processes). Over a wide range of energies, the neutrino-induced event spectrum is very similar in shape to the antineutrino-induced one. However, at energies $E_\nu \sim 6.3$~PeV, around the Glashow resonance ($\bar \nu_e + e^- \rightarrow W^-$), the $\bar \nu_e$-induced event spectrum might present a bump. The spectrum in the PeV range can thus potentially highlight differences between $\nu$ and $\bar \nu$ fluxes.   In practice, the fact that all the events in the HESE sample have energies below the Glashow resonance means that the determination of a possible neutrino-antineutrino symmetry is very challenging, as noted in previous studies~\cite{Barger:2014iua, Palladino:2015uoa}.
 
\item Next, in Sec.~\ref{sec:2powNS}, we have considered a two-component astrophysical neutrino flux with two distinct models. We have studied an isotropic model with two unbroken power-law spectra (Sec.~\ref{sec:2pow}) and concluded that it does not significantly improve the fit to the data over a single power-law flux. Then, we have also considered a model with different spectra and flavor ratios in each hemisphere of origin (Sec.~\ref{sec:NS}), and evaluated the possible existence of a North-South asymmetry. While there is a hint of such an asymmetry in the astrophysical neutrino flux when also using the through-going muon data~\cite{Aartsen:2015knd}, the low upgoing HESE event rate prevents such a determination using only HESE data.
\end{itemize}

Despite only a few tens of observed events, the IceCube detection of extraterrestrial high-energy neutrinos provides a unique window to explore cosmic accelerators and neutrino propagation. The possibility of testing standard and more exotic physics with these neutrinos is very exciting and much work has been devoted to these purposes. As the statistics accumulate, we anticipate a more accurate determination of the flux and thus, of the underlying production mechanisms and propagation properties, as well as the possible connection with other cosmic messengers, as $\gamma$-rays and cosmic-rays. Larger effective volumes and sensitivities of future neutrino observatories~\cite{Aartsen:2014njl, Adrian-Martinez:2016fdl} will be of crucial relevance, bringing us closer to establishing the origin and nature of the high-energy neutrinos in the Universe.

\section*{Acknowledgements}
ACV wishes to thank F. Halzen and C. Arg\"uelles Delgado for their gracious hospitality during his visit to WIPAC. SPR is supported by a Ram\'on y Cajal contract, by the Spanish MINECO under grant FPA2014-54459-P  and by the Generalitat Valenciana under grant PROMETEOII/2014/049. OM is supported by PROMETEOII/2014/050 and by the Spanish grant FPA2014--57816-P of the MINECO. SPR and OM are also supported by the Spanish MINECO grant SEV-2014-0398. The authors are also partially supported by the Marie Sklodowska-Curie grant 674896.  SPR is also partially supported by the Portuguese FCT through the CFTP-FCT Unit 777 (PEst-OE/FIS/UI0777/2013).     

\vspace{-0mm}

\bibliographystyle{apsrev4-1}
\bibliography{IceCube4yr}

\end{document}